\documentclass[10pt,journal,twoside,onecolumn,romanappendices]{IEEEtran}

\usepackage{graphicx}
\usepackage{subfigure}
\usepackage{multirow}
\usepackage{array}
\usepackage{footnote}

\usepackage[ruled,linesnumbered]{algorithm2e}

\usepackage{dsfont}
\usepackage{amsfonts,amsmath}
\usepackage{url,cite}
\usepackage{color}

\def\cred{\textcolor{black}}  

\begin{document}

\title{A Plug-and-Play Priors Framework for Hyperspectral Unmixing}

\author{Min~Zhao,~\IEEEmembership{Student Member,~IEEE},
        Xiuheng~Wang,~\IEEEmembership{Student Member,~IEEE},
        Jie~Chen,~\IEEEmembership{Senior Member,~IEEE}, \\
        Wei~Chen,~\IEEEmembership{Senior Member,~IEEE}
\thanks{Manuscript submitted to IEEE Trans. Geosci. Remote  sens. September 28, 2020; revised November 30, 2020 and accepted
	December 22, 2020.}
\thanks{{A preliminary version of this work has been published
in 2020 IEEE International conference of Image Processing
(ICIP)~\cite{wang2020unmixing}}. The work of J. Chen was supported in part by National Key Research and Development Program of China under Grant 2018AAA0102200 and 111 Project (B18041). The work of W. Chen was supported in part by the Natural Science Foundation of China (61671046) and the Beijing Natural Science Foundation (L202019).  The work of M. Zhao was supported in part  by Innovation Foundation for Doctor Dissertation of Northwestern Polytechnical University (Corresponding author: J. Chen).}
\thanks{M. Zhao and X. Wang contributed equally to this work. M. Zhao, X. Wang and J. Chen are with School of Marine Science and Technology,
Northwestern Polytechnical University, Xi'an 710072, China, and Key Laboratory of Ocean Acoustics and Sensing, Ministry of Industry and Information Technology,
Xi'an 710072, China. W. Chen is with State Key Laboratory of Rail Traffic Control and Safety, Beijing Jiaotong University, China. }
}


\maketitle

\begin{abstract}
Spectral unmixing is a widely used technique in hyperspectral image processing and analysis. It aims to separate mixed pixels into the component materials and their corresponding abundances. Early solutions to spectral unmixing are performed independently on each pixel. Nowadays, investigating proper priors into the unmixing problem has been popular as it can significantly enhance the unmixing performance. However, it is non-trivial to handcraft a powerful regularizer, and complex regularizers may introduce extra difficulties in solving  optimization problems in which they are involved. To address this issue, we present a plug-and-play (PnP) priors framework for hyperspectral unmixing. More specifically, we use the alternating direction method of multipliers (ADMM) to decompose the optimization problem into two
iterative subproblems. One is a regular optimization problem depending on the forward model, and the other is a proximity operator related to the prior model and can be regarded as an image denoising problem. Our framework is flexible and extendable which allows a wide range of denoisers to replace prior models and avoids handcrafting regularizers. Experiments conducted on both synthetic data and real airborne {data} illustrate the superiority of the proposed strategy compared with other state-of-the-art hyperspectral unmixing methods.
\end{abstract}


\IEEEpeerreviewmaketitle

\section{Introduction}
\label{sec:intro} \IEEEPARstart{H}{yperspectral} imaging is a continuously
growing field and has received increasing interest in the past few years. In
contrast to traditional RGB and multispectral images providing a limited
number of spectral channels, hyperspectral images capture hundreds of
spectral bands, and this rich spectral information facilitates the analysis
of elements in the scenes~\cite{stein2002anomaly,chang2004estimation}.
However, due to factors such as the low spatial resolution and the presence
of multiple scattering, an observed pixel usually contains a mixture of
several materials. The presence of mixed pixels restricts the
interpretability of the data and has an adverse impact on the accuracy of
various tasks, such as image classification~\cite{fauvel2012advances} and
target detection\cite{yang2019sparse}. Spectral unmixing is thus playing an
important role in hyperspectral image processing. This technique aims at
decomposing a mixed pixel spectrum into a set of pure spectral signatures
called endmembers, and their corresponding fractional
abundances~\cite{heylen2014review,bioucas2012hyperspectral}.

There exist many hyperspectral unmixing methods, and most of them are based
on the linear mixing model (LMM) due to  its simplicity and physical
interpretability~\cite{dobigeon2009joint,yang2015geometric,yuan2018overview}.
The LMM assumes that an observed pixel spectrum is a linear combination of a
set of endmembers weighted by their associated fractional
abundances~\cite{drumetz2016blind,hong2018augmented}. Though the LMM has
practical advantages, there are many situations where the incoming light may
undergo complex interactions and multiple photon reflections introduce
nonlinear effects on the mixed spectra~\cite{chen2012nonlinear}. The mixing
model is then advantageously formulated as a nonlinear one. \cred{The
work~\cite{hong2018augmented} proposes an augmented LMM to address the
spectral variability issue.} In our work, we focus on designing a flexible
strategy to exploit proper priors to the unmixing task. Without loss of the
generality, we discuss our method based on the LMM.

{The spectral signatures of endmembers can be extracted from
hyperspectral data or selected from a library. Under the assumption that the
endmembers are known in advance, the unmixing task is committed to estimating
the fractional
abundances~\cite{heinz2001fully,chen2012nonlinear,halimi2016fast,ammanouil2015graph}.}
{Early unmixing methods, such as the fully constrained least square
method (FCLS)~\cite{heinz2001fully} and the non-negative constrained least
square error method (NCLS)~\cite{heinz2001fully}, are pixel-by-pixel
algorithms and ignore the prior information existing in hyperspectral
images.}
 {Nevertheless, natural hyperspectral images usually
contain various inherent correlations in both spatial and spectral domains. Properly using this prior information can effectively enhance the
unmixing performance.}

{A significant amount of regularized unmixing algorithms have been proposed
in recent years. One such regularizer is the total-variation (TV)
regularizer. In~\cite{ammanouil2015hyperspectral_1}, a TV regularizer is
imposed on the reconstructed image to incorporate the spatial and spectral
information via pixel links in the unmixing problem. The spatial smoothness
of hyperspectral images can also be employed to abundance maps. The
SUnSAL-TV method~\cite{iordache2012total} introduces a TV regularizer to
enhance the spatial consistency of estimated abundances. However, classical
TV regularizers are restrictive due to their assumption in local spatial
correlations, i.e., a pixel is similar to its
neighbors~\cite{wang2017hyperspectral,he2017total}. The non-local
regularizers take non-local spatial information into consideration to fully
exploit similar patterns and structures across the
image~\cite{feng2016nonlocal,zhong2013non,yao2019nonconvex}. Graph-based
regularizers capture the features in arbitrary
neighborhood~\cite{zhong2013non}. The
works~\cite{wang2017spatial,li2018superpixel} use superpixel methods to
generate the spatial groups and exract local pixels with  homogenous
spectra. In~\cite{Zhang2018Spectral}, $\ell_1$-norm is used to constrain the
spatial and spectral domains to impose sparsity on the solution.}
\cred{In~\cite{hong2018sulora}, subspace unmixing with low-rank attribute
embedding (SULoRA) is proposed to jointly estimate subspace projections and
abundance maps. It solves a constrained optimization problem with subspace
and abundance regularizers.} Therefore, properly {introducing} constraints
and designing ingenious regularizers play an important role in improving the
{unmixing performance}. However, it is a non-trivial task to handcraft an
effective regularizer, and complex regularizers may increase the difficulty
of solving associated optimization problems. \cred{Inspired by the
successful applications of deep learning, convolutional neural networks
(CNNs) have been introduced to learn spatial and spectral priors to conduct
hyperspectral unmixing. The work~\cite{palsson2019convolutional} utilizes a
2D CNN framework to make use of the spatial and spectral information of
hyperspectral images. In~\cite{khajehrayeni2020hyperspectral}
and~\cite{zhang2018hyperspectral}, 3D convolution based models are proposed
to integrate spatial and spectral priors. However, these CNN based methods
only focus on local areas and fail to capture nonlocal features of
hyperspectral images.}

Recently, benefiting from variable splitting techniques, many plug-and-play
priors (PnP) methods have {received} great success in many
hyperspectral image processing tasks {(regarded as linear inverse problems)}, such as
denoising~\cite{lin2019hyperspectral,Zhuang2017Hyperspectral},
deconvolution~\cite{gong2018learning,wang2020learning} and
pansharpening~\cite{teodoro2017sharpening,teodoro2017scene}. Instead of
using handcrafted regularizers, these PnP priors based approaches plug
denoising methods as a module to replace the \emph{proximity operator} in
the iterative optimization
procedure~\cite{sreehari2016plug,teodoro2019image}. The alternating
direction method of multipliers (ADMM)~\cite{boyd2011distributed} is widely
used as the {variable splitting technique} in these image restoration
approaches, for its well-known \cred{convergence
property~\cite{teodoro2017scene,sreehari2016plug}}. {In these methods, the spatial and spectral characteristics existing in
3D domain can be captured with denoisers. Various denoisers, which are regarded  as black-boxes, can be  flexibly injected. Another popular strategy is directly employing deep neural networks to solve linear inverse problems in an end-to-end way, an overview of recent state-of-the-art deep learning methods can be found in~\cite{bai2020deep}.}

In this paper, we propose a novel framework based on ADMM to introduce prior information of abundance maps and hyperspectral images in the unmixing task.
Without handcrafting regularizers, we incorporate different denoisers into this framework to capture various spatial and spectral correlations. Also, adopting denoisers significantly improves the robustness of the proposed unmixing method to noise interruption. The usage of various denoisers demonstrates the flexibility and extendability of our framework. We summarize our contributions as follows:
\begin{itemize}
  \item In our work, we extend the PnP priors framework to propose a
      robust hyperspectral unmixing method. Our scheme is based on ADMM
      algorithm, which decomposes the {associated optimization} problem into two subproblems.
      Specifically, {one of the subproblem is a proximity operator
      related to the prior model and can be replaced by a denoising operator.}
  \item A variety of denoisers are plugged into the proposed unmixing
      framework, including linear and non-linear, band-wise and 3D,
      traditional and deep learning based. These denoising operators allows us to exploit various priors of images and bypass the difficulty in designing regularizers.
  \item The proposed framework is designed to incorporate
      denoisers with two strategies. One exploits the spatial and spectral
      prior information from reconstructed hyperspectral images. {The other one captures the spectral and spatial
      correlations directly from fractional abundance
      maps and alleviates the computational burden.}
\end{itemize}

The paper is organized as follows. In Section~\ref{sec:2}, the outline of
the PnP priors framework for linear inverse problem is presented.
Section~\ref{sec:3} presents the proposed a {unmixing
method based on PnP priors with a variety of denoisers}. In
Section~\ref{sec:4}, both synthetic and real data experiments are conducted
and analyzed. Section~\ref{sec:5} concludes this paper.

\section{Plug-and-play Priors framework}
\label{sec:2} In this section, we outline the PnP priors framework for the
linear inverse problem. Let $\mathbf{d}$ be the observed data, $\mathbf{x}$
is the hidden parameters to be estimated. {A
general description of the linear inverse problem is}
\begin{equation}\label{eq.model}
  \mathbf{d}=\mathcal{S}(\mathbf{x})+\mathbf{n},
\end{equation}
where $\mathcal{S}(\cdot)$ presents the linear forward model, $\mathbf{n}$
is an {additive independent and identically distributed (i.i.d.)} Gaussian noise. {We can estimate $\mathbf{x}$ by seeking the minimum of the following objective function:}
\begin{equation}\label{eq.lms_model}
  \mathbf{\hat{x}}=\arg \min_{\mathbf{x}}\frac{1}{2}\|\mathbf{d}-\mathcal{S}(\mathbf{x})\|_{2}^{2}+\lambda J(\mathbf{x}),
\end{equation}
where $J(\mathbf{x})$ presents a regularizer encoding the prior information
to enforce the desired property of the solution, and $\lambda$ is a positive
penalty parameter to control the impact of {$J(\mathbf{x})$}.

Various algorithms have been proposed to solve this problem, the proposed
framework is based on ADMM to solve the optimization
problem~\eqref{eq.lms_model} owing to its good convergence property. We
introduce an auxiliary variable $\mathbf{v}$ and the
problem~\eqref{eq.lms_model} can be rewritten as:
\begin{equation}\label{eq.lms_model_1}
\begin{split}
  \mathbf{\hat{x}}=\arg \min_{\mathbf{x}}&\frac{1}{2}\|\mathbf{d}-\mathcal{S}(\mathbf{x})\|_{2}^{2}+\lambda J(\mathbf{v}),\\
  &\text{s.t.}~~\mathbf{v}=\mathbf{x}.
  \end{split}
\end{equation}
The associated augmented Lagrangian function is given by
\begin{equation}\label{eq.lms_model_2}
  \mathbf{\hat{x}}=\arg \min_{\mathbf{x}}\frac{1}{2}\|\mathbf{d}-\mathcal{S}(\mathbf{x})\|_{2}^{2}+\lambda J(\mathbf{v})+\frac{\rho}{2}\|\mathbf{x}-\mathbf{v}+\mathbf{u}\|_{2}^{2},
\end{equation}
where $\mathbf{u}$ is the scaled dual variable, and $\rho$ is the penalty
parameter. Then the optimization problem~\eqref{eq.lms_model_2} can be
solved by repeating the following steps:
\begin{align}
   &\tilde{\mathbf{x}} \leftarrow \mathbf{\hat{v}}-\mathbf{u};\\
  \label{eqn1}&\mathbf{\hat{x}} \leftarrow \arg \min_{\mathbf{x}}\frac{1}{2}\|\mathbf{d}-\mathcal{S}(\mathbf{x})\|_{2}^{2}
  +\frac{\rho}{2}\|\mathbf{\tilde{x}}-\mathbf{x}\|_{2}^{2}; \\
  &\tilde{\mathbf{v}}\leftarrow \mathbf{\hat{x}}+\mathbf{u};\\
  \label{eqn2}&\mathbf{\hat{v}}\leftarrow \arg \min_{\mathbf{v}}\lambda J(\mathbf{v})
  +\frac{\rho}{2}\|\mathbf{\tilde{v}}-\mathbf{v}\|_{2}^{2};\\
  \label{eqn3}&\mathbf{u}\leftarrow \mathbf{u}+\mathbf{\hat{x}}-\mathbf{\hat{v}}.
\end{align}
The steps include two key operators. The step~\eqref{eqn1} is a linear
inverse operator, and the step~\eqref{eqn2} is the proximity operator that
can be considered as a denoising problem. {From a Bayesian viewpoint, the problem in step~\eqref{eqn2} can be regarded as a denoising problem. We can replace this step with a variety of denoisers rather than
solving it with an explicit form.} This strategy avoids handcrafting the
regularizer $J(\mathbf{v})$ and exploits priors of the hidden parameters by
using denoisers.

\section{Application to spectral unmixing}
\label{sec:3}
\subsection{Problem Formulation}
We denote a hyperspectral image as
$\mathbf{Y}_{\text{3D}}\in\mathbb{R}^{L\times M\times K}$, where $L$, $M$
and $K$ are the numbers of spectral bands, rows and columns of the image.
For ease of mathematical formulation, the columns of image
$\mathbf{Y}_{\text{3D}}$ are stacked to form
$\mathbf{Y}\in\mathbb{R}^{L\times N}$ with $N=M\times K$ representing the
number of hyperspectral pixels. The LMM for a hyperspectral image is based
on the assumption that a pixel of hyperspectral data is {a linear
mixture of endmembers.} We express it as:
\begin{equation}\label{eq.LMM}
  \mathbf{Y}=\mathbf{MA}+\mathbf{N},
\end{equation}
where $\mathbf{Y}=[\mathbf{y}_1,\cdots,\mathbf{y}_{N}]\in\mathbb{R}^{L\times
N}$ and $\mathbf{y}_i$ is the signature vector corresponding to the $i$-th
pixel in the hyperspectral image. $\mathbf{M}\in\mathbb{R}^{L\times P}$ is
the endmember matrix containing $P$ spectral signatures.
$\mathbf{A}=[\mathbf{a}_1,\cdots,\mathbf{a}_{N}]\in\mathbb{R}^{P\times N}$
is the abundance matrix and $\mathbf{a}_i$ denotes the corresponding
fractional abundances of the $i$-th hyperspectral pixel.
$\mathbf{N}=[\mathbf{n}_1,\cdots,\mathbf{n}_{N}]\in\mathbb{R}^{L\times N}$
represents the {i.i.d.}
Gaussian noise. As the abundances represent the relative amount of each
endmember, they need to be non-negative {(abundance nonnegativity constraint, ANC).} Furthermore, the entire
spectra are decomposed into endmember contributions, and thus the abundance
fractions should satisfy {the abundance sum-to-one constraint (ASC)}. The ANC and ASC can be defined as follows:
\begin{align}
  \label{eq.ANC}&{a}_{ij}\geq 0,   \quad \forall i,j\\
  \label{eq.ASC}&\sum_{i=1}^{P}a_{ij}=1,\quad \forall j.
\end{align}
With a known endmember matrix $\mathbf{M}$, the estimation of abundances can
be obtained by solving the following optimization problem,
\begin{equation}
\begin{split}
\label{eq.lms_model_1}&\min_{\mathbf{A}}\frac{1}{2}\|\mathbf{Y}-\mathbf{MA}\|_{\text{F}}^{2},\\
  \text{s.t.}~~{a}_{ij}&\geq 0, \quad\forall i,j, \quad\sum_{i=1}^{P}{a}_{ij}=1, \quad\forall j,
\end{split}
\end{equation}
where $\|\cdot\|_\text{F}$ is the Frobenius norm of the matrix. Compared to
performing unmixing on each individual pixel, {incorporating proper regularizers is effective to} enhance the unmixing performance.
The regularization term can be imposed on the reconstructed image or directly on
the abundances.
In our work, we {aim to exploit priors of reconstructed images and
abundance maps respectively} according to the different choices of the pattern switch matrix
$\mathbf{H}$, and the optimization problem in~\eqref{eq.lms_model_1} can be
formulated as:
\begin{equation}
\begin{split}\label{eq.lms_model_22}
  &\min_{\mathbf{A}}\frac{1}{2}\|\mathbf{Y}-\mathbf{MA}\|_{\text{F}}^{2}+\lambda\Phi(\mathbf{HA})\\
  \text{s.t.}~~&{a}_{ij}\geq 0, \quad\forall i,j, \quad\sum_{i=1}^{P}a_{ij}=1, \quad\forall j.
\end{split}
\end{equation}
{The squared-error term $\|\mathbf{Y}-\mathbf{MA}\|_{\text{F}}^{2}$ is  the data fidelity term while the second
term $\Phi(\mathbf{HA})$ is a regularization term, which is defined to further promote desired
property of the results. The proposed framework is suggested to be conducted using two forms:
\begin{itemize}
	\item  Pro-H: $\mathbf{H}=\mathbf{M}$, $\Phi(\mathbf{HA})$ is used to penalize on reconstructed images;
    \item Pro-A: $\mathbf{H}=\mathbf{I}$ where $\mathbf{I}$ is identity matrix, $\Phi(\mathbf{HA})$ is used to penalize on abundance maps directly.
\end{itemize}
$\lambda$ is a positive parameter
trading off the data fidelity term and the regularization term.}
\subsection{Proposed Plug-and-Play Framework for Hyperspectral Unmixing}
{Designing a powerful regularizer is a non-trivial task, and complex
regularizers usually make the optimization problem more complicated to
solve. In our work, we aim to use the variable splitting strategy to create
a flexible and extendable PnP priors framework to tackle the unmixing
problem. It allows various image denoisers (linear or non-linear, band-wise
or 3D, traditional or deep learning based) to replace the design of
regularizers. More specially, the optimization
problem~\eqref{eq.lms_model_22} is decoupled into two sub-problems: a
quadratic programming (QP) sub-problem with linear constraints and \cred{an}
image denoising sub-problem. These sub-problems are then solved iteratively
to estimate the final abundance.}

{ADMM is adopted to decouple the data fidelity term and the regularization term in~\eqref{eq.lms_model_22}.} By introducing an
auxiliary variable $\mathbf{Z}$, problem~\eqref{eq.lms_model_22} is equivalent to:
\begin{equation}
	\begin{split}\label{eq.object_equ}
  &\min_{\mathbf{A},\mathbf{Z}}\frac{1}{2} \|\mathbf{Y}-\mathbf{MA}\|_{\text{F}}^{2}+\lambda\Phi(\mathbf{Z})\\
  &\text{s.t.}  ~~\mathbf{Z}=\mathbf{HA}\\
  &~~~~~{a}_{ij}\geq 0, \quad\forall i,j, \quad\sum_{i=1}^{P}a_{ij}=1, \quad\forall j\cred{.}
\end{split}
\end{equation}
The corresponding augmented Lagrangian function with the scaled dual variable $\mathbf{U}$ is
\begin{equation}\label{eq.Lag}
\begin{aligned}
\mathcal{L}_{\rho}({\bf A,Z}) = & \arg\mathop{\min}_{{\bf A, \bf Z}}\, \frac{1}{2}\|\mathbf{Y}-\mathbf{MA}\|_{\text{F}}^{2} + \lambda \Phi ({\bf Z}) \\\
&+\frac{\rho}{2}\|{\bf HA-Z+U}\|_{\text{F}}^{2}\\
  \text{s.t.}~~{a}_{ij}&\geq 0, \quad\forall i,j, \quad\sum_{i=1}^{P}a_{ij}=1, \quad\forall j
\end{aligned}
\end{equation}
where $\rho$ is a positive penalty parameter. {According} to the PnP priors
framework for the linear inverse problem,~\eqref{eq.Lag} can be solved by
repeating the following steps until convergence:
\begin{equation}\label{eq.stepa}
	\begin{aligned}
	 {\bf A}_{k+1} =& \arg\mathop{\min}_{{\bf A}}\,\frac{1}{2}\|{\bf Y}-{\bf M}{\bf A}_k\|_{\text{F}}
	 +\frac{\rho_k}{2}\|{\bf HA}_k-{\bf\tilde{\bf X}}_{k}\|_{\text{F}}\\
	 \text{s.t.}&~~{a}_{ij}\geq 0, \quad\forall i,j, \quad\sum_{i=1}^{P}a_{ij}=1, \quad\forall j.
\end{aligned}
\end{equation}
\begin{equation}\label{eq.stepb}
\begin{aligned}
		&{\bf Z}_{k+1} = \arg\mathop{\min}_{{\bf Z}}\,\lambda \Phi ({\bf Z})+\frac{\rho_k}{2}\|{\tilde{\bf Z}}_{k}-{\bf Z}\|_{\text{F}}^{2}\\
\end{aligned}
\end{equation}
\begin{equation}\label{eq:stepc}
\begin{aligned}
	&{\bf U}_{k+1} = {\bf U}_k+{\bf HA}_{k+1}-{\bf Z}_{k+1}
\end{aligned}
\end{equation}
\begin{equation}\label{eq:stepd}
\begin{aligned}
&{\rho}_{k+1} = {\alpha \rho}_{k}
\end{aligned}
\end{equation}
where ${\tilde{\bf X}}_{k} = {\bf Z}_{k} - {\bf U}_{k}$ and ${\tilde{\bf
Z}}_{k} = {\bf HA}_{k+1} +{\bf U}_k$. The objective
function~\eqref{eq.object_equ} is thus decoupled into two
subproblems~\eqref{eq.stepa} and~\eqref{eq.stepb}. We solve the QP
problem~\eqref{eq.stepa} pixel by pixel. For a hyperspectral pixel ${\bf
y}_i$, this problem can be reformulated as
\begin{equation}\label{eq.stepa_equ}
\begin{aligned}
{\bf a}_{k+1, i}& = \arg\mathop{\min}_{{\bf a}_i}\,\frac{1}{2}\|{\bf y}_i-{\bf M}{\bf a}_i\|^{2}+\frac{\rho_k}{2}\|{\bf Ha}_i-{\bf\tilde{\bf x}}_{k, i}\|^{2}\\
&= \arg\mathop{\min}_{{\bf a}_i} \frac{1}{2}{{\bf a}_i^{\top}}{\bf Q}{\bf a}_i + {\bf f}^{\top}{\bf a}_i\\
&\text{s.t.}~~{a}_{ij}\geq 0, \quad\forall i,j, \quad\sum_{i=1}^{P}a_{ij}=1, \quad\forall j\cred{,}
\end{aligned}
\end{equation}
where $\mathbf{Q} =
\mathbf{M}^{\top}\mathbf{M}+{\rho_k}\mathbf{H}^{\top}\mathbf{H}$ and
$\mathbf{f} = -(\mathbf{M}^{\top}\mathbf{y}_i
+{\rho_k}\mathbf{H}^{\top}\tilde{\mathbf{x}}_{k,i})$. The QP problem
in~\eqref{eq.stepa_equ} is a standard FCLS problem, which can be solved by a
generic QP solver. The second step involving regularizer in~\eqref{eq.stepb}
can be rewritten as:
\begin{equation}\label{eq:stepb_equ}
{\bf Z}_{k+1} = \arg\mathop{\min}_{{\bf Z}}\,\frac{1}{2(\sqrt{{\lambda}/{\rho_k}})^2}\|{\tilde{\bf Z}}_{k}-{\bf Z}\|_{\text{F}}^{2}+\Phi ({\bf Z}).
\end{equation}

{Based on Bayesian theory, the problem in~\eqref{eq:stepb_equ} can be
regarded as \cred{an} image denoising problem to remove} additive Gaussian
noise with a standard deviation $\sigma_{n}=\sqrt{{\lambda}/{\rho_k}}$.
 {In form Pro-H, this step aims to obtain a clean reconstructed hyperspectral image ${\bf Z}_{k+1}$ from the noisy observations $\tilde {\bf Z}_{k}$ with a noise level
$\sqrt{{\lambda}/{\rho_k}}$ . In form Pro-A, this step becomes an
abundance maps denoising operator. It is worth noting that, compared to Pro-H, Pro-A alleviates the computational burden since abundance maps always have smaller volumes than reconstructed hyperspectral images.}

Instead of solving the step~\eqref{eq.stepb} {in an explicit form, we replace it with} various denoisers. In other words, the denoiser extracting prior
information is plugged into the iterative algorithm to solve the
step~\eqref{eq.stepb}, which is so-called PnP priors. The denoising operator
is actually performed in the 3D domain, we rewrite~\eqref{eq:stepb_equ} as
follows:
\begin{equation}\label{eq:denosier}
{\bf Z}_{k+1} = \mathcal{T}^{-1}(\textit{Denoiser}({\mathcal{T}(\tilde{\bf Z}}_{k}), \sqrt{{\lambda}/{\rho_k}})),
\end{equation}
where $\mathcal{T}(\cdot)$ is an operator to reshape a 2D matrix to a 3D
data cube {and $\mathcal{T}^{-1}(\cdot)$ is the opposite operator.} In this way, the input of the denoiser is a 3D image cube and
the denoiser can jointly capture the spatial and spectral information.

In order to force the constraint $\mathbf{Z}=\mathbf{HA}$ to be ideal and
affect the noise level $\sqrt{{\lambda}/{\rho}}$ to make the denosier more
conservative with iterations, we increase the penalty parameter $\rho$ with
scaling factor $\alpha$ during the iterations in~\eqref{eq:stepd}. The
procedure of our PnP priors framework is summarized in Algorithm
\ref{alg_1}.

\begin{algorithm}[!t]
	\label{alg_1}
	\KwData{A hyperspectral image $\bf Y$, endmembers $\bf M$, the number of hyperspectral pixels $N$, the regularization parameter $\lambda$, the penalty factor $\rho$, the number of iterations $K$, the scaling factor $\alpha$, the pattern switch matrix $\bf H$.}
	\KwResult{Fractional abundance $\bf A$.}
	Initialize ${\bf A}$ randomly, initialize the auxiliary variable ${\bf Z}_0={\bf MA}_0$,\\ scaled dual variable ${\bf U}_0=\bf 0$, $k=0$\;
	\While{Stopping criterias are not met and $k\le K$}{
		${\tilde{\bf X}}_{k} = {\bf Z}_k - {\bf U}_k$\;
		\For{$i=1$ to $N$}{
			Calculate ${\bf a}_{k+1,i}$ in~\eqref{eq.stepa_equ} by using a generic QP solver\;
		}{
			${\tilde{\bf Z}}_{k} = {\bf HA}_{k+1} + {\bf U}_k$\;
			${\bf Z}_{k+1} = \mathcal{T}^{-1}(\textit{Denoiser}(\mathcal{T}({\tilde{\bf Z}}_{k}), \sqrt{{\lambda}/{\rho_k}}))$\;
			${\bf U}_{k+1} = {\bf U}_k + {\bf HA}_{k+1} - {\bf Z}_{k+1}$\;
			$\rho_{k+1} = \alpha\rho_{k}$\;
			$k = k + 1$\;
		}
	}
	\caption{Plug-and-play priors framework for hyperspectral unmixing.}
\end{algorithm}
\subsection{Denoisers Plugged into Plug-and-Play Priors Framework}
{Benefiting from the generalization ability of the proposed PnP priors framework for
hyperspectral unmixing, different denoising algorithms can be plugged into it as
prior models.} Our framework is thus opening a huge opportunity to adopt
a wide variety of prior information. In our work, we use popular and
effective denoising operators in the PnP framework iterations, including
linear or non-linear, band-wise or 3D, and traditional or deep learning
based denoisers. Specifically, this work investigates the following denoisers:
\begin{itemize}
  \item \textbf{Non-local means denoising (NLM)}~\cite{buades2011non}: NLM
      is a filter based image deniosing algorithm. This denoiser is linear
      and based on a non-local averaging of {similar} pixels in the image.
      Our proposed unmixing methods plugging NLM are named as
      Pro-\cred{H}-NLM and Pro-A-NLM, respectively.
  \item \textbf{Block-matching and 3D filtering
      (BM3D)}~\cite{dabov2007image}: BM3D is a non-linear denoiser
      proposed to use collaborative filter to reduce the noise of grouped
      image blocks. We denote our proposed unmixing methods using the BM3D
      as Pro-\cred{H}-BM3D and Pro-A-BM3D, respectively.
  \item \textbf{BM4D}~\cite{maggioni2012nonlocal}: BM4D is a 3D cube based
      denoising algorithm extended from BM3D, which has been successfully
      applied to volumetric images. It can make full use of the spectral
      information as well as spatial correlation of volumetric data. We
      denote our proposed unmixing methods using BM4D denoiser as
      Pro-\cred{H}-BM4D and Pro-A-BM4D, respectively.
  \item \textbf{A total variation regularized low-rank tensor
      decomposition denoising model
      (LRTDTV)}~\cite{wang2017hyperspectral_1}: LRTDTV is a hyperspectral
      denoising method using an anisotropic spatial-spectral total
      variation regularization to characterize the smooth priors from
      spatial and spectral domains. The tensor Tucker decomposition is
      used to describe the global correlation in all bands. We use this
      method to denoise reconstructed hyperspectral images and name the
      corresponding unmixing method as Pro-\cred{H}-LRTDTV.
  \item \textbf{Denoising convolutional neural network
      (DnCNN)}~\cite{zhang2017beyond}: DnCNN is a deep learning based
      denoiser using a convolutional neural network. We use this method to
      denoise the abundance maps and this strategy is named as
     Pro-A-DnCNN.
     Considering the similar text
     features between natural images and abundance maps, we directly use
     this denoising network with default network parameters trained by
     natural image datasets, which is online
     available\footnote{https://github.com/cszn/DnCNN}. \cred{Besides, a
     transfer learning strategy is considered, we generate the ground
     truth of abundance maps to fine tune models trained by natural
     images.}
\end{itemize}

\section{Experiments}
\label{sec:4} In this section, we illustrate the performance of the proposed
PnP priors based unmixing methods. Experiments on both synthetic data and
real airborne images are conducted. Several
state-of-the-art methods, FCLS~\cite{heinz2001fully},
SUnSAL-TV~\cite{iordache2012total}, CsUnL0~\cite{shi2018collaborative} and
SCHU~\cite{li2018superpixel} are implemented to compare with the proposed
unmixing methods. FCLS is a conventional unmixing method {and} aims to
minimize the least square error with the ANC and ASC. SUnSAL-TV is a sparse
unmixing method with a TV-norm regularizer {and solved using ADMM.}
CsUnL0 is a collaborative sparse unmixing approach using $\ell_{0}$-norm.
SCHU is a graph-based unmixing method via the superpixel, {which divides pixels of an image into local clusters that exhibit similar features.}

The performance of abundance estimation is evaluated using the root-mean-square-error (RMSE) given by
\begin{equation}\label{eq.rmse}
  \text{RMSE} = \sqrt{\frac{1}{NP}\sum_{i=1}^{N}\|\mathbf{a}_{i}-\hat{\mathbf{a}}_{i}\|^{2}},
\end{equation}
where $\mathbf{a}_{i}$ and $\hat{\mathbf{a}}_{i}$ denote the true and
estimated abundance vectors of the $i$-th pixel, respectively. Further, we
use the peak signal-to-noise ratio (PSNR) to evaluate the image denoising
quality, which is defined by:
\begin{equation}\label{eq.PSNR}
  \text{PSNR} = 10\times \log_{10}\left(\frac{\text{MAX}^{2}}{\text{MSE}}\right),
\end{equation}
\cred{we use the reconstructed hyperspectral image defined by
$\mathbf{\hat{Y}}=\mathbf{M}\hat{\mathbf{A}}$ and the ground truth
$\mathbf{\tilde{Y}}=\mathbf{M}\mathbf{A}$ to calculate PSNR. $\text{MAX}$ is
the maximum pixel value of $\mathbf{\hat{Y}}$, and MSE is the mean square
error between $\mathbf{\hat{Y}}$ and $\mathbf{\tilde{Y}}$, which is defined
by:
\begin{equation}
  \text{MSE}=\frac{1}{MK}\sum_{i=1}^{M}\sum_{j=1}^{K}\|\mathbf{\hat{Y}}(i,j)-\mathbf{\tilde{Y}}(i,j)\|^{2},
\end{equation}
where $M$ and $K$ are the numbers of rows and columns of the image.}

\subsection{Experiments with Synthetic Dataset}
In these experiments, we generate a synthetic dataset to evaluate the
unmixing methods in both quantitative and qualitative manners.
\subsubsection{Data Description}
Synthetic data are generated using the U.S. Geological Survey (USGS)
spectral library. The USGS spectral library consists of 500 mineral
materials, which can be download at the
website\footnote{http://speclab.cr.usgs.gov/spectral.lib06}. The
hyperspectral image consists of 224 contiguous bands. We use $4$ spectral
signatures ($P=4$) which are randomly selected from the USGS spectral
library to build the endmember matrix $\mathbf{M}$. This data is generated
following the method of Hyperspectral Imagery Synthesis
tools\footnote{http://www.ehu.es/ccwintco/index.php/Hyperspectral Imagery
Synthesis tools for MATLAB}, and we use Gaussian Fields to form the
abundance matrix $\mathbf{A}$. \cred{There are both mixed and pure pixels in
the dataset, and our dataset is also built with spatial correlation between
neighboring pixels.} The abundances are restricted to satisfy the ANC and
ASC. The spatial resolution is set to $256\times 256$, and all spectral
bands are used in our synthetic data. In order to verify the denoising
ability of the proposed method, we add zero-mean Gaussian noise to this
data, with the SNR setting to 5~dB, 10~dB, 20~dB and 30~dB, respectively.
The ground-truth reference of synthetic data is shown in the first column of
Figure~\ref{fig:map_syn_abu}.

\cred{We generate 400 simulated abundance maps with the spatial resolution
$256\times 256$ to train the models of DnCNN. The learning rate is set to
one hundredth of the original one, and the batch size is set to 16. We fine
tune the network with different SNRs, namely, 10~dB, 15~dB, 20~dB, 25~dB,
30~dB, 35~dB, 40~dB, 45~dB, 50~dB, 55~dB and 60~dB.}
\subsubsection{Unmixing Results and Discussion}
\begin{figure}[!t]
\centering
\includegraphics[width=9cm]{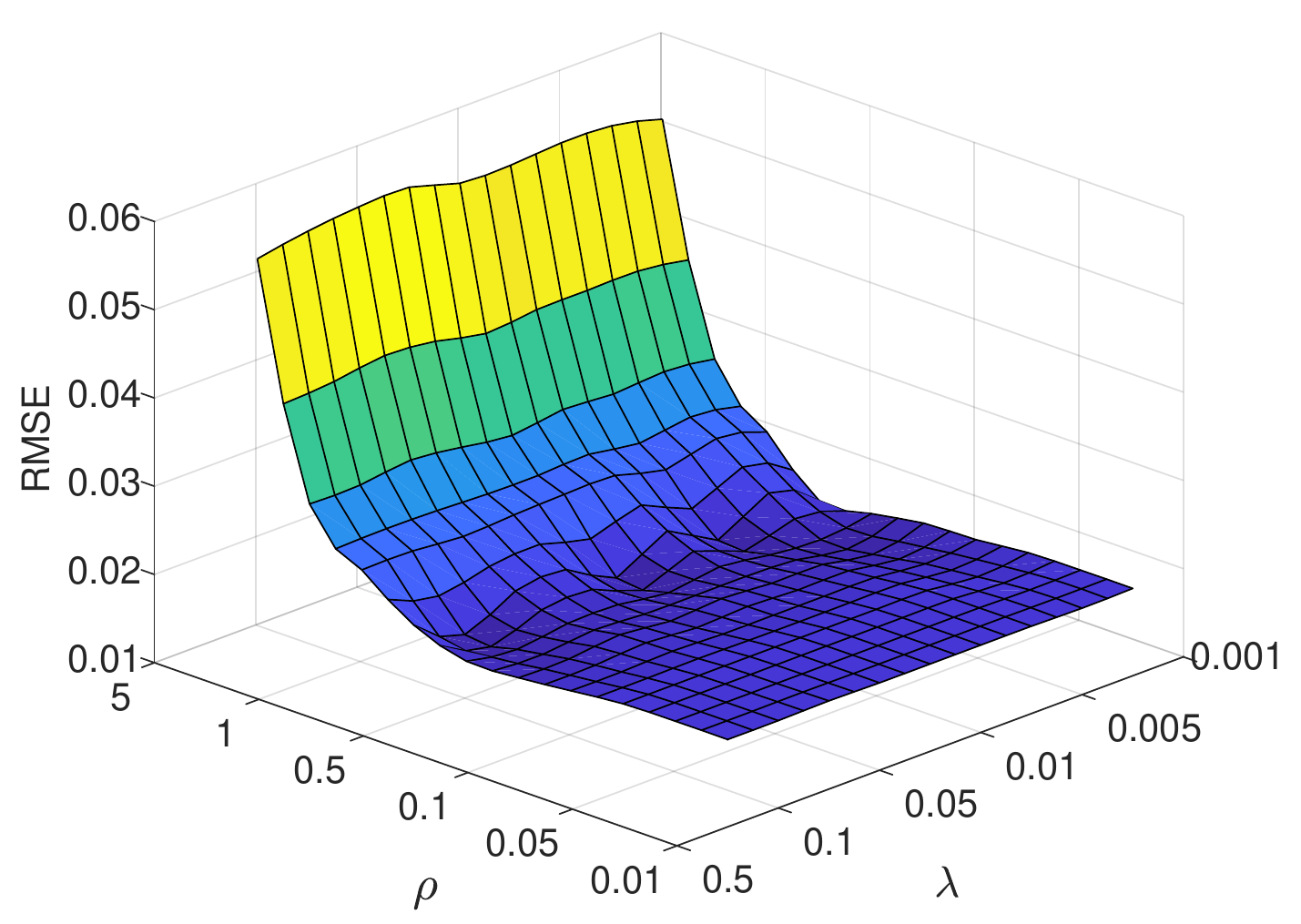}
\caption{RMSE as a function of the regularization parameters for the Pro-\cred{H}-NLM with synthetic data (SNR = 20 dB).}
\label{fig:loss}
\end{figure}

The setting values of $\rho$ and $\lambda$ used in experiments of this
dataset are presented in Table~\ref{tab:result_RMSE_PSNR}. We set $\alpha=1$
for experiments of denoising hyperspectral images ({Pro-H}), and
$\alpha=1.1$ for experiments of denoising abundance maps ({Pro-A}).
Figure~\ref{fig:loss} illustrates how parameters $\rho$ and $\lambda$ affect
the performance of Pro-\cred{H}-NLM using the synthetic data with SNR=20~dB.
We can see that within a reasonable range of parameter values our method
obtains a {satisfactory} unmixing result, and a good selection of $\rho$ and
$\lambda$ makes the method easily find a solution near the optimal result.
It is noted that $\lambda$ controls the {impact of regularization term and
has an effect on the unmixing results. $\rho$ is a penalty parameter in the
augmented Lagrangian function and only influence the convergence speed.}

\begin{table*}[!htp]
\footnotesize \centering
\caption{\small RMSE and PSNR Comparison of Synthetic Dataset.}
\renewcommand\arraystretch{1.5}
\begin{tabular}{c|c|c|c|c|cc|c|c}
\hline
\hline
                             & \multicolumn{2}{c|}{5dB}                                                                          & \multicolumn{2}{c|}{10dB}                                                                                                & \multicolumn{2}{c|}{20dB}                                                                         & \multicolumn{2}{c}{30dB}                                                                         \\ \hline
                             & RMSE                                             & PSNR                                           & RMSE                                                         & PSNR                                                      & \multicolumn{1}{c|}{RMSE}                                  & PSNR                                 & RMSE                                            & PSNR                                            \\ \hline
FCLS                         & 0.0897                                           & 31.194                                         & 0.0581                                                       & 35.435                                                    & \multicolumn{1}{c|}{0.0200}                                & 44.964                               & 0.0064                                          & 55.647                                          \\ \hline
\multirow{2}{*}{SUnSAL-TV}   & 0.0802                                           & 31.278                                         & 0.0492                                                       & 35.698                                                    & \multicolumn{1}{c|}{0.0199}                                & 45.892                               & 0.0064                                          & 55.774                                          \\ \cline{2-9}

                             & \multicolumn{2}{c|}{($\lambda_{\text{TV}}=1\times 10^{-2}$)} & \multicolumn{2}{c|}{($\lambda_{\text{TV}}=5\times 10^{-3}$)}                        & \multicolumn{2}{c|}{($\lambda_{\text{TV}}=5\times 10^{-3}$)} & \multicolumn{2}{c}{($\lambda_{\text{TV}}=5\times 10^{-3}$)} \\ \hline
\multirow{2}{*}{CsUnL0}      & 0.1078                                           & 30.819                                         & 0.0695                                                       & 35.199                                                    & \multicolumn{1}{c|}{0.0242}                                & 44.734                               & 0.0077                                          & 54.709                                          \\ \cline{2-9}
                             & \multicolumn{2}{c|}{($\lambda_{\text{rs}}=1\times 10^{-3}$)}                                      & \multicolumn{2}{c|}{($\lambda_{\text{rs}}=1\times 10^{-2}$)}                                                             & \multicolumn{2}{c|}{($\lambda_{\text{rs}}=1\times 10^{-4}$)}                                       & \multicolumn{2}{c}{($\lambda_{\text{rs}}=5\times 10^{-4}$)}                                      \\ \hline
\multirow{2}{*}{SCHU}        & 0.1078                                           & 30.822                                         & 0.0695                                                       & 35.205                                                    & \multicolumn{1}{c|}{0.0242}                                & 44.740                               & 0.0077                                          & 54.711                                          \\ \cline{2-9}
                            & \multicolumn{2}{c|}{($\mu=0.001, \lambda_{\text{s}}=5\times 10^{-3}$)}
                            & \multicolumn{2}{c|}{($\mu=0.005
                            , \lambda_{\text{s}}=1\times 10^{-2}$)}                                         & \multicolumn{2}{c|}{($\mu=0.003
                            , \lambda_{\text{s}}=1\times 10^{-3}$)}                  & \multicolumn{2}{c}{($\mu=0.0001
                            , \lambda_{\text{s}}=1\times 10^{-2}$)}                  \\ \hline

\multirow{2}{*}{Pro-\cred{H}-NLM}                    & 0.0615
& 32.601                                         & 0.0418
& 37.351                                                    &
\multicolumn{1}{c|}{\textbf{0.0172}}                                & 46.335
& 0.0062                                          & 55.710
\\ \cline{2-9}
                             & \multicolumn{2}{c|}{($\rho=1, \lambda=3\times 10^{-3}$)}                                          & \multicolumn{2}{c|}{($\rho=0.5, \lambda=1\times 10^{-3}$)}                                                               & \multicolumn{2}{c|}{($\rho=0.1, \lambda=2\times 10^{-4}$)}                                      & \multicolumn{2}{c}{($\rho=0.005, \lambda=1\times 10^{-4}$)}                                      \\ \hline
\multirow{2}{*}{Pro-\cred{H}-BM3D}  & 0.0612 & {33.557} & \textbf{0.0391} &
\textbf{38.489} & \multicolumn{1}{c|}{0.0181} & 46.147 & 0.0062 & 55.813
\\ \cline{2-9}
                             & \multicolumn{2}{c|}{($\rho=0.1, \lambda=5\times 10^{-3}$)}                                        & \multicolumn{2}{c|}{($\rho=0.5, \lambda=5\times 10^{-3}$)}                                                               & \multicolumn{2}{c|}{($\rho=0.01, \lambda=1\times 10^{-3}$)}                                       & \multicolumn{2}{c}{($\rho=0.001, \lambda=2\times 10^{-4}$)}                                      \\ \hline
\multirow{2}{*}{Pro-\cred{H}-BM4D}  & 0.0809
& 33.077                                         & 0.0537
& 37.009                                                    &
\multicolumn{1}{c|}{0.0190}                                & \textbf{46.338}
& \textbf{0.0061}                                          & \textbf{56.142}
\\ \cline{2-9}
                             & \multicolumn{2}{c|}{($\rho=1, \lambda=1\times 10^{-3}$)}                                          & \multicolumn{2}{c|}{($\rho=5, \lambda=2\times 10^{-4}$)}                                                                 & \multicolumn{2}{c|}{($\rho=0.01, \lambda=5\times 10^{-4}$)}                                       & \multicolumn{2}{c}{($\rho=0.005, \lambda=5\times 10^{-4}$)}                                      \\ \hline
\multirow{2}{*}{Pro-\cred{H}-LRTDTV}  & 0.0718
& 32.226                                         & 0.0502
& 37.099                                                    &
\multicolumn{1}{c|}{0.0193}                                & 46.072
& 0.0063                                          & 55.968
\\ \cline{2-9}
                             & \multicolumn{2}{c|}{($\rho=0.01, \lambda=5\times 10^{-4}$)}                                          & \multicolumn{2}{c|}{($\rho=0.05, \lambda=2\times 10^{-4}$)}                                                                 & \multicolumn{2}{c|}{($\rho=0.01, \lambda=5\times 10^{-3}$)}                                       & \multicolumn{2}{c}{($\rho=0.01, \lambda=5\times 10^{-3}$)}                                      \\ \hline
\multirow{2}{*}{Pro-A-NLM}   & 0.0760                                           & 32.723                                         & 0.00471                                                      & 37.490                                                    & \multicolumn{1}{c|}{0.0184}                                & 46.419                               & 0.0062                                          & 56.034                                          \\ \cline{2-9}
                             & \multicolumn{2}{c|}{($\rho=3, \lambda=5\times 10^{-5}$)}                                          & \multicolumn{2}{c|}{($\rho=2, \lambda=1\times 10^{-5}$)}                                                                 & \multicolumn{2}{c|}{($\rho=5, \lambda=3\times 10^{-4}$)}                                          & \multicolumn{2}{c}{($\rho=5, \lambda=1\times 10^{-4}$)}                                          \\ \hline
\multirow{2}{*}{Pro-A-BM3D}  & 0.0900                                           & 31.777                                         & 0.0581                                                       & 36.310                                                    & \multicolumn{1}{c|}{0.0199}                                & 45.958                               & 0.0063                                          & 55.950                                          \\ \cline{2-9}
                             & \multicolumn{2}{c|}{($\rho=4, \lambda=1\times 10^{-3}$)}                                          & \multicolumn{2}{c|}{($\rho=10, \lambda=1\times 10^{-4}$)} & \multicolumn{2}{c|}{($\rho=5, \lambda=2\times 10^{-4}$)}                                          & \multicolumn{2}{c}{($\rho=8, \lambda=5\times 10^{-5}$)}                                          \\ \hline
\multirow{2}{*}{Pro-A-BM4D}  & 0.0890                                           & 31.742                                         & 0.0574                                                       & 36.293                                                    & \multicolumn{1}{c|}{0.0200}                                & 45.934                               & 0.0064                                          & 55.916                                          \\ \cline{2-9}
                             & \multicolumn{2}{c|}{($\rho=5, \lambda=2\times 10^{-5}$)}                                          & \multicolumn{2}{c|}{($\rho=10, \lambda=1\times 10^{-4}$)}                                                                & \multicolumn{2}{c|}{($\rho=8, \lambda=2\times 10^{-4}$)}                                          & \multicolumn{2}{c}{($\rho=10, \lambda=1\times 10^{-4}$)}                                         \\ \hline
\multirow{2}{*}{Pro-A-DnCNN} & \textbf{\cred{0.0603}}&\textbf{\cred{33.807}}
& \cred{0.0472} & \cred{37.449} & \multicolumn{1}{c|}{\cred{0.0191}}
&\cred{46.181} &\cred{\textbf{0.0061}}& \cred{55.981}
\\ \cline{2-9}
                             & \multicolumn{2}{c|}{(\cred{$\rho=8, \lambda=1\times 10^{-4}$})}  & \multicolumn{2}{c|}{(\cred{$\rho=5, \lambda=1\times 10^{-4}$})}   & \multicolumn{2}{c|}{(\cred{$\rho=8, \lambda=2\times 10^{-4}$})}   & \multicolumn{2}{c}{(\cred{$\rho=8, \lambda=1\times 10^{-4}$})}\\ \hline\hline
\end{tabular}
\vspace{1mm}
\\\footnotesize{Boldface numbers are the lowest RMSEs and the highest PSNRs.}\\
  \label{tab:result_RMSE_PSNR}
\end{table*}

\begin{figure*}[!t]
\centering
\includegraphics[width=18.5cm]{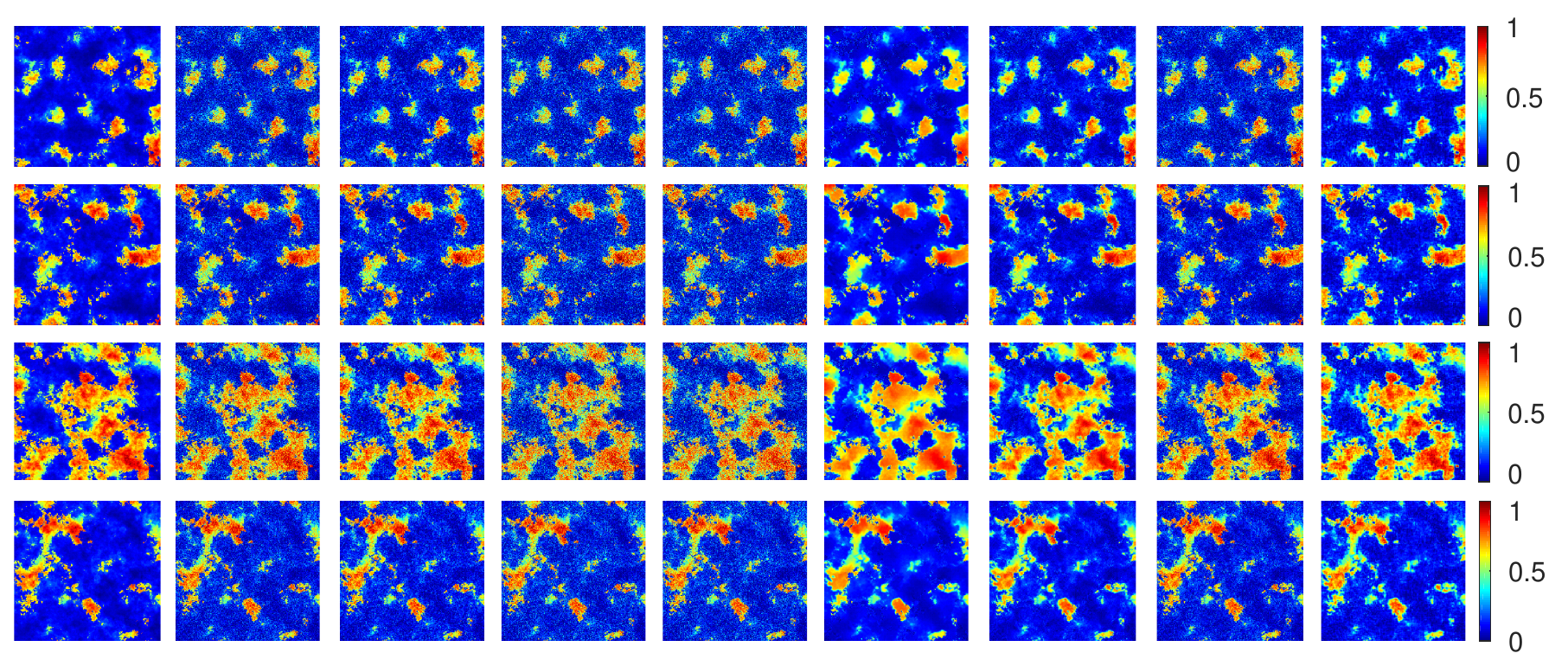}
\caption{Abundance maps of synthetic data (SNR=5dB). From top to bottom: different
endmembers. From left to right: ground-truth, FCLS, SUnSAL-TV, CsUnL0, SCHU, Pro-\cred{H}-NLM, Pro-\cred{H}-BM3D, Pro-\cred{H}-BM4D, and Pro-\cred{H}-LRTDTV.}
\label{fig:map_syn_abu}
\end{figure*}

\begin{figure*}[!t]
\centering
\includegraphics[width=18.5cm]{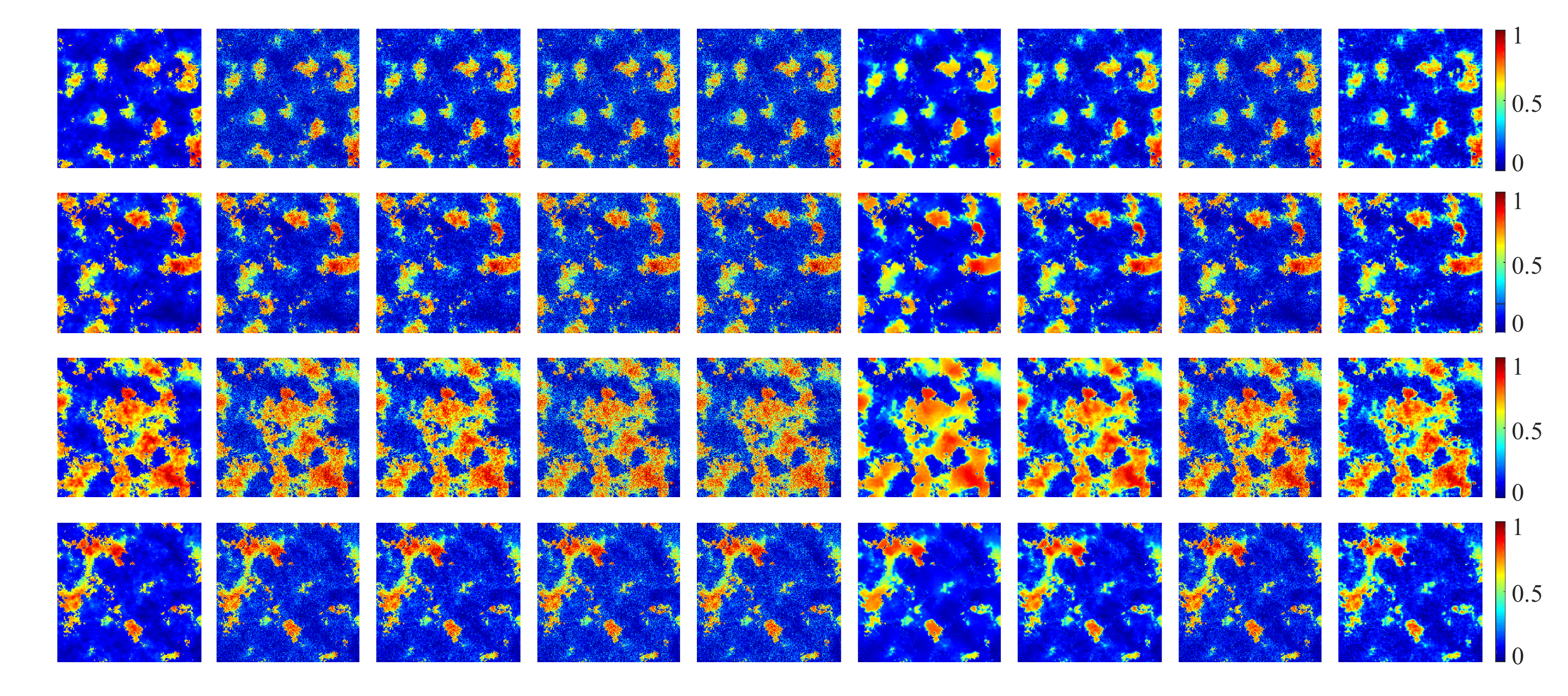}
\caption{\cred{Abundance maps of synthetic data (SNR=10dB). From top to bottom: different
endmembers. From left to right: ground-truth, FCLS, SUnSAL-TV, CsUnL0, SCHU, Pro-A-NLM, Pro-A-BM3D, Pro-A-BM4D, and Pro-A-DnCNN.}}
\label{fig:map_syn_abu_1}
\end{figure*}

Table~\ref{tab:result_RMSE_PSNR} shows the RMSE results of FCLS, SUnSAL-TV,
CsUnL0, SCHU and our proposed PnP priors based unmixing methods. As we can
see, for the synthetic data, RMSE results of the proposed framework show
superiority compared to the other methods, especially when data SNRs are
low. This indicates the effectiveness of priors excavated by denoisers.
Whereas FCLS is a pixel-based unmixing method and it ignores the
spatial-spectral correlations of natural hyperspectral images; SUnSAL-TV
considers the spatial smooth information in the optimization problem.
However, this method only considers the similarity of the four neighbors of
a pixel. On the other hand, it neglects the variability information across
spectral bands. CsUnL0 has no regard for the spatial information in the
hyperspectral image, and SCHU does not introduce the spectral priors. All
these compared methods use fixed regularizers and lack flexibility, while
the proposed framwork does not need to handcraft regularizers and models the
priors with the help of various denoisers. We can also observe that the
unmixing results of denoisers plugged in the reconstructed image are better
than denoisers plugged in the abundance maps. In other words, the features
captured form the reconstructed image can effectively enhance the unmixing
performance compared with the features captured form the abundance maps.
Figure~\ref{fig:map_syn_abu} presents the abundance maps of four compared
methods, Pro-\cred{H}-NLM, Pro-\cred{H}-BM3D, Pro-\cred{H}-BM4D and
Pro-\cred{H}-LRTDTV, with 5~dB data. In Figure~\ref{fig:map_syn_abu_1}, we
present the abundance maps of four compared methods, Pro-A-NLM, Pro-A-BM3D,
Pro-A-BM4D and Pro-A-DnCNN, with 10~dB data. We can observe that all the
estimated abundance maps are consistent with the ground-truth. However, the
abundance maps of our methods are with less noise and more close to
ground-truth.

\begin{figure*}[!t]
\centering
\includegraphics[width=18.5cm]{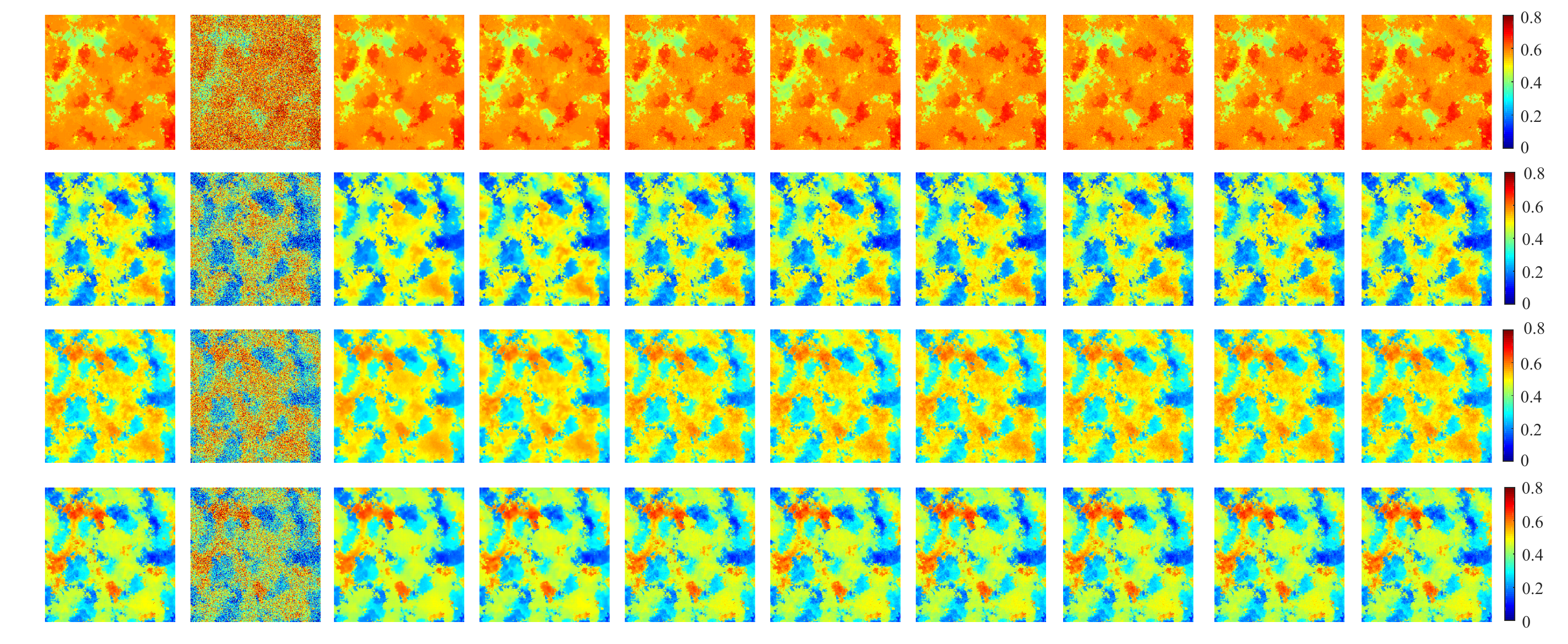}
\caption{\cred{The reconstructed image maps. From top to bottom: reconstructed images
in different channels (50, 100, 150, 200). From left to right: clean image, noisy image,
reconstructed images of Pro-\cred{H}-NLM, Pro-\cred{H}-BM3D, Pro-\cred{H}-BM4D, Pro-\cred{H}-LRTDTV, Pro-A-NLM, Pro-A-BM3D, Pro-A-BM4D, and Pro-A-DnCNN.}}
\label{fig:map_syn_image}
\end{figure*}

In addition, the PSNR results are also shown in
Table~\ref{tab:result_RMSE_PSNR} to demonstrate the impact of denoisers and
show the superiority of our proposed framework. {These results show that our
proposed methods enhance the unmixing performance with priors form clean latent hyperspectral images by the application of denoisers.} In Figure~\ref{fig:map_syn_image}, we present the image reconstructed by the abundances and endmembers with LMM of different bands using the synthetic
data with 10~dB. It is clear that our reconstructed images are cleaner
compared with original noisy data and closer to the clean data. Using
denoisers allows our method to maintain good unmixing performance when SNR
is low. When SNR is high, the unmixing performance is still satisfactory,
but the effect of denoisers is slight. The synthetic experiments confirm the
applicability of our proposed methods from both quantitative and qualitative
aspects.

\subsubsection{Convergence Analysis} It is well known that a standard ADMM
algorithm is guaranteed to
converge~\cite{eckstein1992douglas,afonso2010fast}. However, in our work, we
use a denoiser as a black-box to replace the explicitly solving the second
subproblem~\eqref{eq.stepb}, and thus the convergence of this operator is
difficult to analysis. In~\cite{sreehari2016plug}, the convergence of the
PnP priors approach based on the linear filter, such as NLM, has been
proved. For non-linear denoisers involving amounts of complex operators
(e.g., BM3D and DnCNN), it is hard to prove the convergence of the proposed
method theoretically. In practice, we observe the convergence curves of our
proposed ADMM-based unmixing scheme involving both linear and non-linear
denoisers to show its good convergence property.

\begin{figure*}[htbp]
\centering
\subfigure{
\includegraphics[width=7cm]{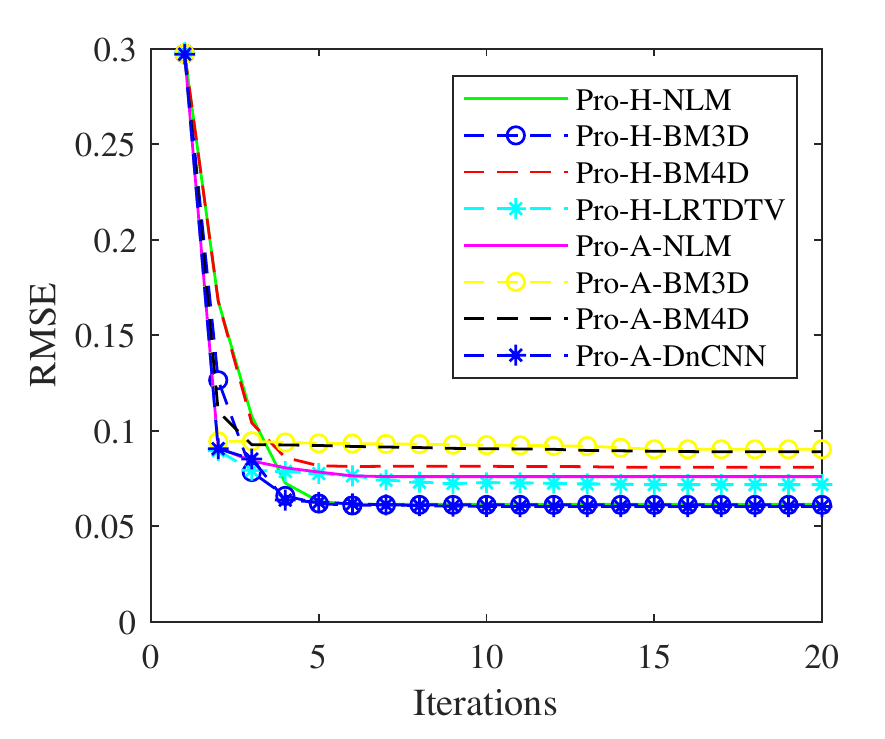}
}
\subfigure{
\includegraphics[width=7cm]{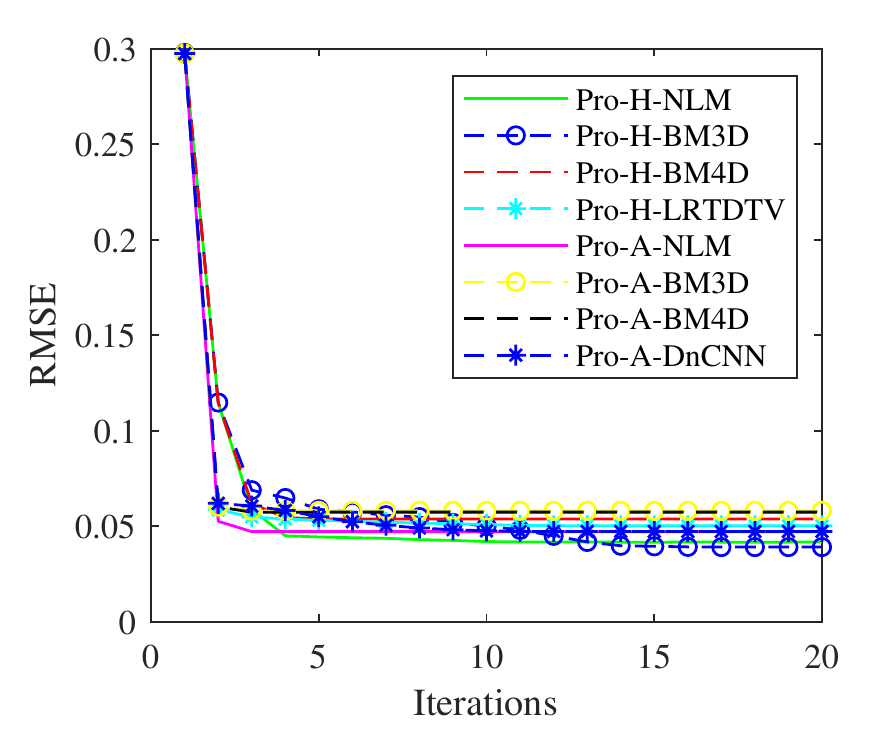}
}
\subfigure{
\includegraphics[width=7cm]{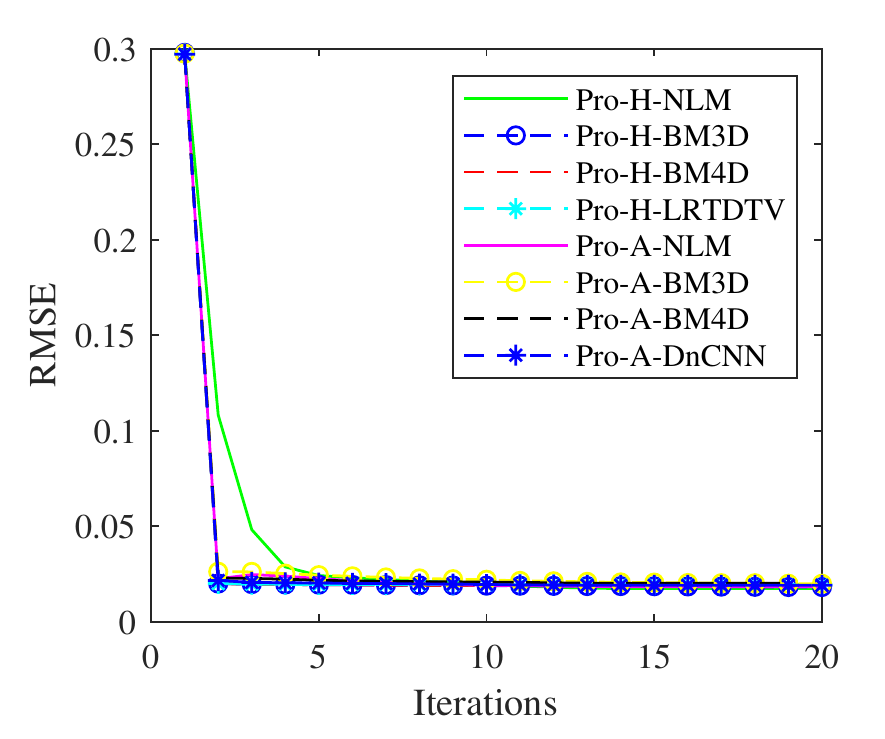}
}
\subfigure{
\includegraphics[width=7cm]{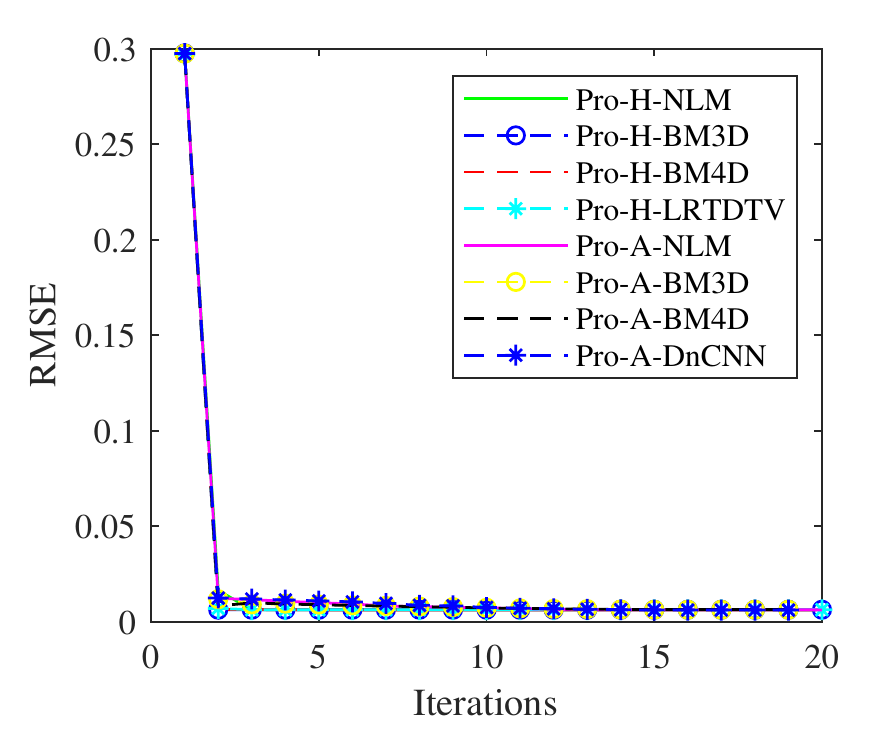}
}
\caption{\cred{The RMSE convergence curves of synthetic data of our proposed methods
 with different SNRs. From left to right, top to bottom: SNR=5~dB, 10~dB, 20~dB and 30~dB, respectively.}}
\label{fig:loss_curve}
\end{figure*}

Figure~\ref{fig:loss_curve} illustrates the RMSE convergence curves of
synthetic data of our proposed unmixing methods \cred{with different SNRs}.
It can be observed that the proposed ADMM-based framework with each
denoising algorithm (both linear and non-linear) has a stable and robust
convergence property. Moreover, the proposed method with all the deniosers
achieve a low RMSE after 3 iterations, which indicates that an early stop
operator can be considered for less computation time.
\subsection{Experiments with Real Hyperspectral Dataset}
Previous experiments were based on synthetic data, which can avoid extra errors caused by atmospheric or camera effects. However, hyperspectral
unmixing methods are designed to be applied directly on real hyperspectral
images. In this subsection, we use two real data to further verify the
effectiveness of our proposed methods.
\begin{figure*}[!t]
\centering
\includegraphics[width=18.5cm]{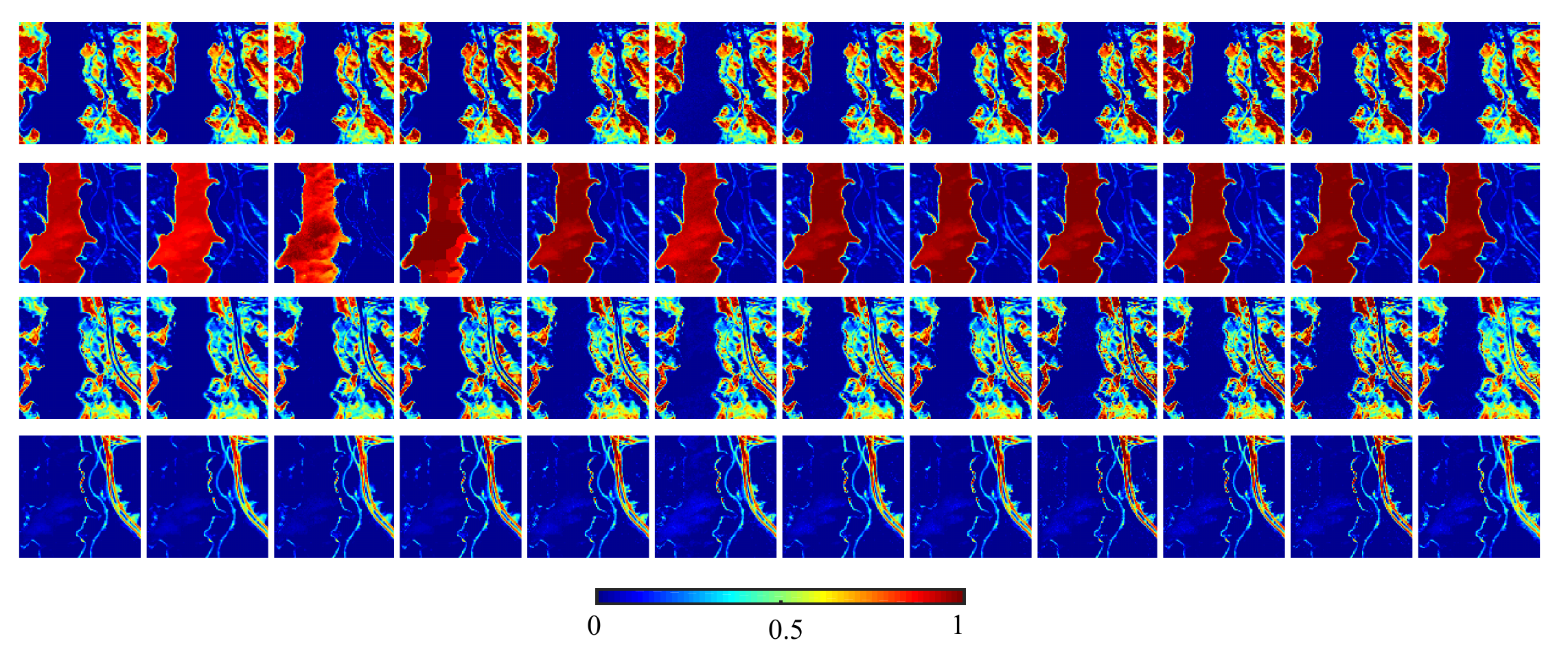}
\caption{Abundance maps of Jasper Ridge dataset. From top to bottom: four endmembers,
 tree, water, soil, and road. From left to right: FCLS, SUnSAL-TV, CsUnL0, SCHU,
 Pro-\cred{H}-NLM, Pro-\cred{H}-BM3D, Pro-\cred{H}-BM4D, Pro-\cred{H}-LRTDTV, Pro-A-NLM, Pro-A-BM3D, Pro-A-BM4D, and Pro-A-DnCNN.}
\label{fig:map_jas}
\end{figure*}

\subsubsection{Data Description}
\begin{figure}[!t]
\centering
\includegraphics[width=8cm]{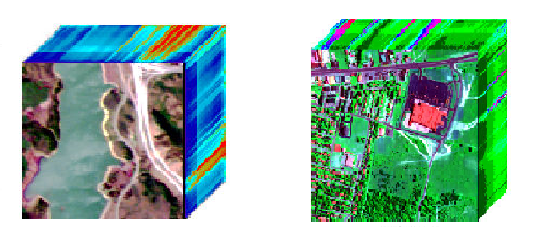}
\caption{Color image of real dataset, Jasper Ridge dataset (left), Urban dataset (right).}
\label{fig:image_real}
\end{figure}

The first dataset is Jasper Ridge dataset which was obtained by the Airborne
Visible Infrared Imaging Spectrometer (AVIRIS). The original size of this
data is $512\times 614$, with $224$ spectral bands. Following the previous
unmixing works\cite{aggarwal2016hyperspectral}, we removed the noisy and
water vapor absorption bands (1-3, 108-112, 154-166 and 220-224) of this
data with 198 exploitable bands remained. An interest subset of $100\times
100$ size was selected to evaluate the unmixing performance. The color image
of this dataset is shown in the left of Figure~\ref{fig:image_real}. There
are four main endmembers in this dataset, including ``tree", ``water",
``soil" and ``road". The endmembers used for this experiment were download
from the website\footnote{https://rslab.ut.ac.ir/data}.

The second real hyperspectral dataset used for real hyperspectral image
experiments is Urban dataset. The Urban dataset is of size $307\times 307$.
The original Urban dataset consists of 210 bands, ranging from 400 nm to
2500 nm. We removed the water absorption and noisy channels (1-4, 76, 87,
101-111, 136-153, 198-210) in our experiment with 162 exploitable bands
remained. Six endmembers, including ``asphalt", ``grass", ``tree", ``roof",
``metal" and ``dirt'', are extracted for this experiment. The endmembers
used for this experiment was also download from the website$^{\text{3}}$.
The color image of this dataset is shown in the right of
Figure~\ref{fig:image_real}.

\cred{For real hyperspectral dataset, it is impossible to obtain a dataset
with the ground-truth of abundances to train the supervised denoising method
DnCNN. We directly use the denoising network with parameters trained by
natural image datasets.}

\begin{figure*}[!t]
\centering
\includegraphics[width=18.5cm]{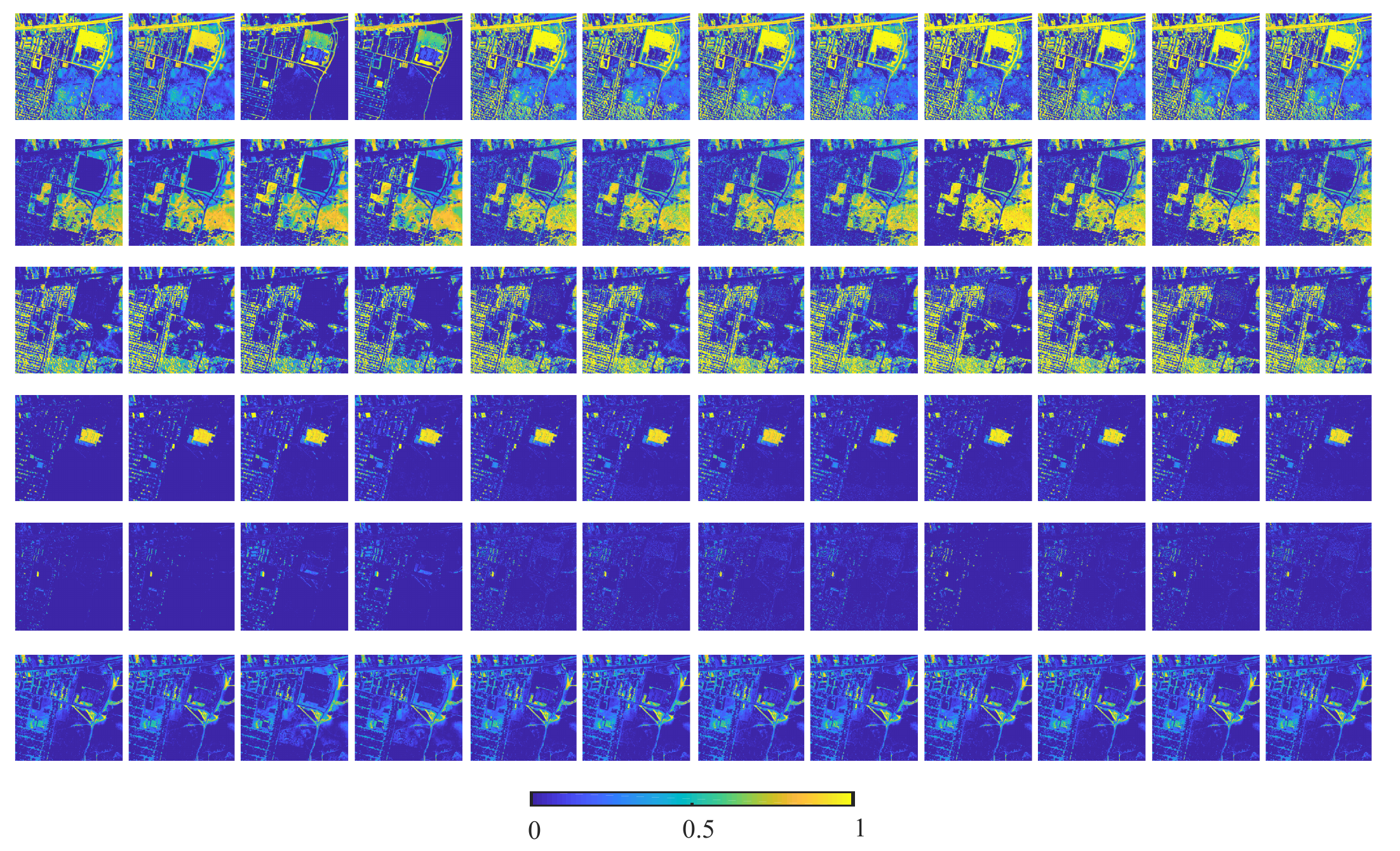}
\caption{Abundance maps of Urban dataset. From top to bottom: six endmembers, asphalt, grass, tree, roof, metal,
and dirt. From left to right: FCLS, SUnSAL-TV, CsUnL0, SCHU, Pro-\cred{H}-NLM, Pro-\cred{H}-BM3D, Pro-\cred{H}-BM4D, Pro-\cred{H}-LRTDTV,
 Pro-A-NLM, Pro-A-BM3D, Pro-A-BM4D, and Pro-A-DnCNN. }
\label{fig:map_urban}
\end{figure*}

\subsubsection{Unmixing Results and Discussion}
The parameter settings of these experiments are the same as the synthetic
data. Figure~\ref{fig:map_jas} presents the abundance maps of different
methods of the Jasper Ridge dataset. The unmixing results of our proposed
methods show the most smooth and clear abundance maps, especially the
unmixing results of the material ``water". For the material ``road", our
methods can emphasize some locations and provide sharper maps. The abundance
maps of the Urban dataset are shown in Figure~\ref{fig:map_urban}. This
dataset is more complex and increases the difficulty of unmixing. A visual
comparison of these methods indicates that our methods preserve more spatial
homogeneous information and contain more details. From
Figure~\ref{fig:map_urban}, we see that our methods provide clear and
smooth abundance maps while CsUnL0 and SCHU fail to unmix this
dataset with the material ``asphalt". All unmixing results of this two real datasets
confirm that our PnP priors based framework for hyperspectral unmixing
provides superior unmixing performance.

\begin{figure*}[t]
  \centering
  \includegraphics[width=16cm]{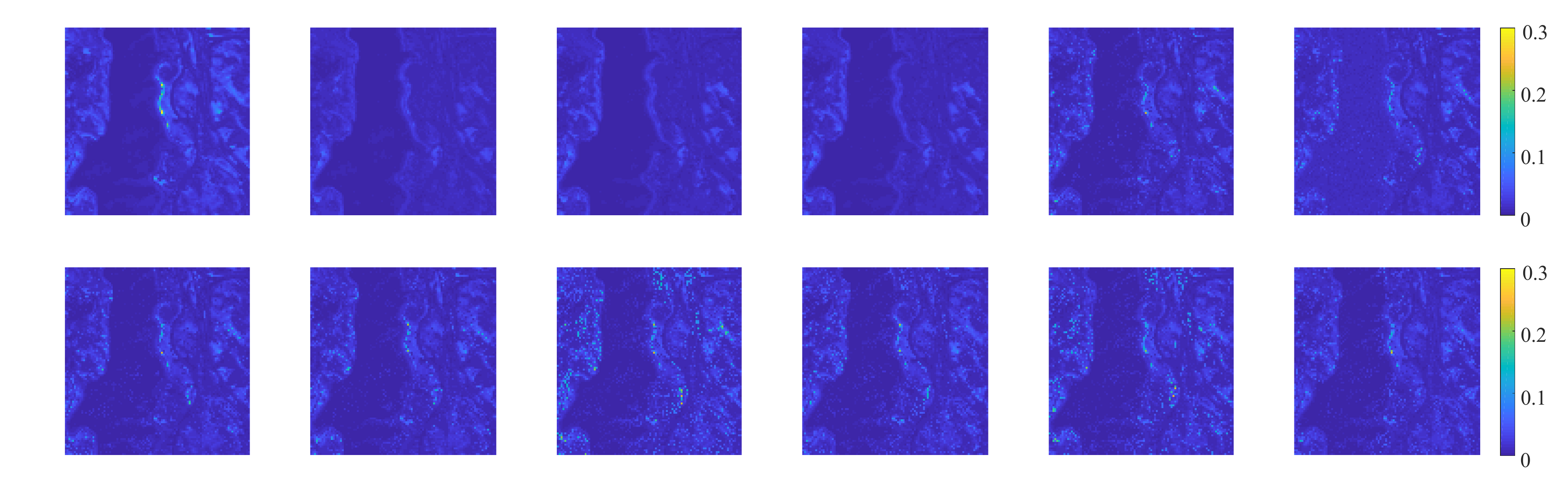}\\
  \caption{\cred{Maps of reconstructed error of Jasper Ridge dataset. From left to right, top to bottom: FCLS, SUnSAL-TV,
   CsUnL0, SCHU, Pro-H-NLM, Pro-H-BM3D, Pro-H-BM4D, Pro-H-LRTDTV, Pro-A-NLM, Pro-A-BM3D, Pro-A-BM4D and Pro-A-DnCNN.}}
  \label{fig_RE_jas}
\end{figure*}

\begin{figure*}[t]
  \centering
  \includegraphics[width=16cm]{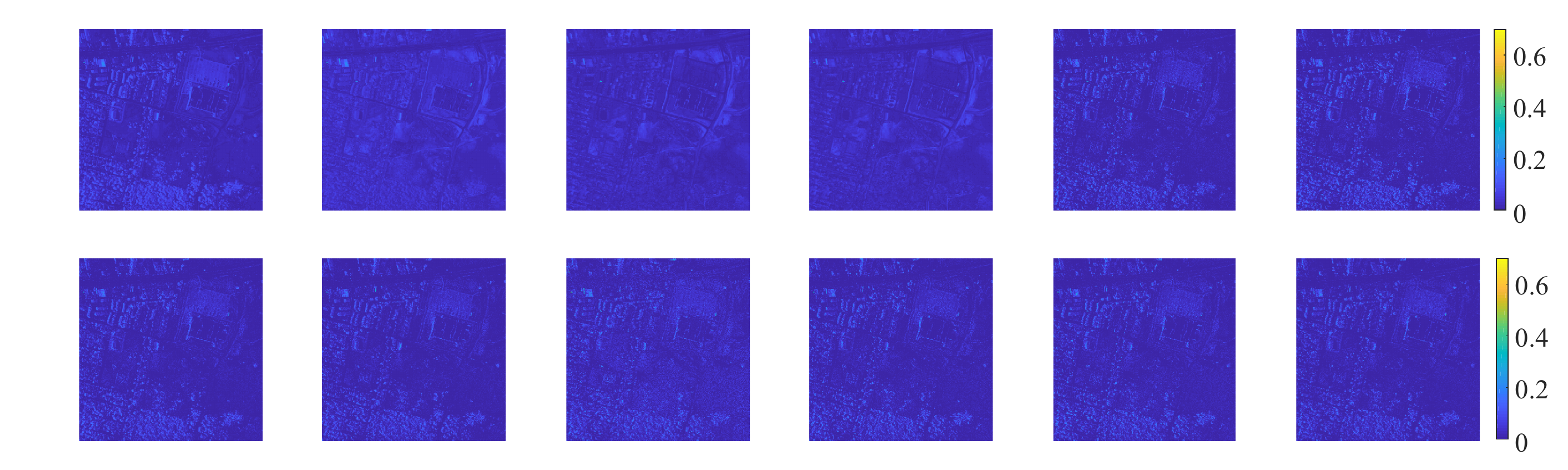}\\
  \vspace{-3pt}
  \caption{\cred{Maps of reconstructed error of Urban dataset. From left to right, top to bottom: FCLS, SUnSAL-TV,
   CsUnL0, SCHU, Pro-H-NLM, Pro-H-BM3D, Pro-H-BM4D, Pro-H-LRTDTV, Pro-A-NLM, Pro-A-BM3D, Pro-A-BM4D and Pro-A-DnCNN.}}
  \label{fig_RE_urban}
\end{figure*}

\cred{We use the reconstructed error (RE) to give a quantitative evaluation
of the unmixing results of real data. RE is defined as:
\begin{equation}
\text{RE}=\sqrt{\frac{1}{NL}\sum_{i=1}^{N}\|\mathbf{y}_i-\mathbf{\hat{y}}_i\|^{2}},
\end{equation}
where $\mathbf{y}_i$ is the $i$-th pixel and $\mathbf{\hat{y}}_i$ is its
reconstruction. Note that the RE results are not necessarily proportional to
the quality of the abundance estimation, and therefore it can only be
considered as complementary information. Table~\ref{tab:RE_jas} shows the RE
results of Jasper Ridge dataset. Table~\ref{tab:RE_urban} presents the RE
results of Urban dataset. We can see that the RE results are all in an order
of magnitude. As the noise are unavoidable in real data, and the proposed
framework can effectively denoise the data, the RE results of the proposed
methods are easily larger than compared methods. The maps of reconstructed
error of Jasper Ridge dataset and Urban dataset are shown in
Figure~\ref{fig_RE_jas} and Figure~\ref{fig_RE_urban}, respectively.}
\begin{table*}[t]
\footnotesize \centering
\caption{\small \cred{RE Comparison of Jasper Ridge dataset.}}
\renewcommand{\arraystretch}{1.15}
\cred{
\begin{tabular}{ccccccc}
\hline
\hline
   & FCLS       & SUnSAL-TV    & CsUnL0    & SCHU       & Pro-H-NLM  & Pro-H-BM3D  \\
RE & 0.0281     & 0.0157       & 0.0166    & 0.0166     & 0.0301     & 0.0312      \\ \hline
   & Pro-H-BM4D & Pro-H-LRTDTV & Pro-A-NLM & Pro-A-BM3D & Pro-A-BM4D & Pro-A-DnCNN \\
RE & 0.0296     & 0.0300       & 0.0363    & 0.0315     & 0.0367     & 0.0304      \\
\hline\hline
\end{tabular}   }
\label{tab:RE_jas}
\end{table*}

\begin{table*}[t]
\footnotesize \centering
\caption{\small \cred{RE Comparison of Urban dataset.}}
\renewcommand{\arraystretch}{1.15}
\cred{
\begin{tabular}{ccccccc}
\hline\hline
   & FCLS       & SUnSAL-TV    & CsUnL0    & SCHU       & Pro-H-NLM  & Pro-H-BM3D  \\
RE & 0.0411     & 0.0183       & 0.0144    & 0.0145     & 0.0555     & 0.0550      \\ \hline
   & Pro-H-BM4D & Pro-H-LRTDTV & Pro-A-NLM & Pro-A-BM3D & Pro-A-BM4D & Pro-A-DnCNN \\
RE & 0.0559     & 0.0540       & 0.0514    & 0.0493     & 0.0529     & 0.0540      \\ \hline\hline
\end{tabular}     }
\label{tab:RE_urban}
\end{table*}

\subsection{\cred{Running Time}}
\cred{To weigh the trade-off between performance gain and computational
burden, we conduct experiments for evaluating the running time using
different datasets. We set the number of iteration $K = 20$. Note that all
the experiments are conducted on the computer with Intel i7-8700 3.2-Ghz CPU
and 16-GB random access memory. Table~\ref{tab:time_1},
Table~\ref{tab:time_2} and Table~\ref{tab:time_3} present the average
computing time of synthetic data, Jasper Ridge dataset and Urban dataset. We
observe that the running time of the proposed framework mainly depends on
the complexity of denoisers. The convergence speed of 2D denoisers is faster
than 3D counterparts. Denoising abundance maps is faster than denoising
hyperspectral images.}
\begin{table}[t]
\footnotesize \centering
\caption{\small \cred{Time consuming of the proposed methods for synthetic data (seconds).}}
\renewcommand{\arraystretch}{1.15}
\cred{
\begin{tabular}{ccccc}
\hline\hline
        & Pro-H-NLM & Pro-H-BM3D & Pro-H-BM4D & Pro-H-LRTDTV \\
time & 288      &  374       &6314        &3247          \\ \hline
        & Pro-A-NLM & Pro-A-BM3D & Pro-A-BM4D & Pro-A-DnCNN  \\
time & 23        & 25         & 47        & 771         \\ \hline\hline
\end{tabular}  }
\label{tab:time_1}
\end{table}

\begin{table}[t]
\footnotesize \centering
\caption{\small \cred{Time consuming of the proposed methods for Jasper Ridge dataset (seconds).}}
\renewcommand{\arraystretch}{1.15}
\cred{
\begin{tabular}{ccccc}
\hline\hline
        & Pro-H-NLM & Pro-H-BM3D & Pro-H-BM4D & Pro-H-LRTDTV \\
time & 48      &  53       &1497        &439          \\ \hline
        & Pro-A-NLM & Pro-A-BM3D & Pro-A-BM4D & Pro-A-DnCNN  \\
time & 4        & 4         & 8        & 192         \\ \hline\hline
\end{tabular} }
\label{tab:time_2}
\end{table}

\begin{table}[h]
\footnotesize \centering
\caption{\small \cred{Time consuming of the proposed methods for Urban dataset (seconds).}}
\renewcommand{\arraystretch}{1.15}
\cred{
\begin{tabular}{ccccc}
\hline\hline
        & Pro-H-NLM & Pro-H-BM3D & Pro-H-BM4D & Pro-H-LRTDTV \\
time & 315      &  403       &5879        &3399          \\ \hline
        & Pro-A-NLM & Pro-A-BM3D & Pro-A-BM4D & Pro-A-DnCNN  \\
time & 37        & 44         & 191        & 1531         \\ \hline\hline
\end{tabular} }
\label{tab:time_3}
\end{table}
\section{Conclusion}
\label{sec:5} In this paper, we propose a PnP priors based
framework for hyperspectral unmixing, which is flexible and extendable by using various image denoisers. The proposed framework makes full use of the spatial-spectral priors of
hyperspectral images and abundance maps. It is easy to
switch the pattern of imposing the penalty on reconstructed hyperspectral images or abundance maps by
different choices of matrix $\mathbf{H}$. Various
denoisers, including linear, non-linear, or deep learning based denoisers are used to excavate the priors. Our proposed unmixing framework with all these
denoisers have a good convergence property. Experiment results of both
synthetic data and real data show that our framework achieves superior
unmixing performance. Also, it indicates the effectiveness of excavated priors via denoisers. In future work, we will focus on how to automatically select the parameters $\rho$ and $\lambda$. Moreover, we will conduct the PnP priors
framework for jointly deblurring and unmixing hyperspectral images.

\bibliographystyle{IEEEtran}
\bibliography{IEEEfull,BIB}\ 

\begin{thebibliography}{10}
\providecommand{\url}[1]{#1}
\csname url@samestyle\endcsname
\providecommand{\newblock}{\relax}
\providecommand{\bibinfo}[2]{#2}
\providecommand{\BIBentrySTDinterwordspacing}{\spaceskip=0pt\relax}
\providecommand{\BIBentryALTinterwordstretchfactor}{4}
\providecommand{\BIBentryALTinterwordspacing}{\spaceskip=\fontdimen2\font plus
\BIBentryALTinterwordstretchfactor\fontdimen3\font minus
  \fontdimen4\font\relax}
\providecommand{\BIBforeignlanguage}[2]{{%
\expandafter\ifx\csname l@#1\endcsname\relax
\typeout{** WARNING: IEEEtran.bst: No hyphenation pattern has been}%
\typeout{** loaded for the language `#1'. Using the pattern for}%
\typeout{** the default language instead.}%
\else
\language=\csname l@#1\endcsname
\fi
#2}}
\providecommand{\BIBdecl}{\relax}
\BIBdecl

\bibitem{wang2020unmixing}
X.~Wang, M.~Zhao, and J.~Chen, ``Hyperspectral unmixing via plug-and-play
  priors,'' in \emph{2020 IEEE International Conference on Image Processing
  (ICIP)}.\hskip 1em plus 0.5em minus 0.4em\relax IEEE, 2020, pp. 1063--1067.

\bibitem{stein2002anomaly}
D.~W. Stein, S.~G. Beaven, L.~E. Hoff, E.~M. Winter, A.~P. Schaum, and A.~D.
  Stocker, ``Anomaly detection from hyperspectral imagery,'' \emph{IEEE Signal
  Proc. Mag.}, vol.~19, no.~1, pp. 58--69, 2002.

\bibitem{chang2004estimation}
C.-I. Chang and Q.~Du, ``Estimation of number of spectrally distinct signal
  sources in hyperspectral imagery,'' \emph{IEEE Trans. Geosci. Remote Sens.},
  vol.~42, no.~3, pp. 608--619, 2004.

\bibitem{fauvel2012advances}
M.~Fauvel, Y.~Tarabalka, J.~A. Benediktsson, J.~Chanussot, and J.~C. Tilton,
  ``Advances in spectral-spatial classification of hyperspectral images,''
  \emph{Proc. of the IEEE}, vol. 101, no.~3, pp. 652--675, 2012.

\bibitem{yang2019sparse}
X.~Yang, J.~Chen, and Z.~He, ``Sparse-spatialcem for hyperspectral target
  detection,'' \emph{IEEE J. Sel. Top. Appl. Earth Observat. Remote Sens.},
  vol.~12, no.~7, pp. 2184--2195, 2019.

\bibitem{heylen2014review}
R.~Heylen, M.~Parente, and P.~Gader, ``A review of nonlinear hyperspectral
  unmixing methods,'' \emph{IEEE J. Sel. Top. Appl. Earth Observat. Remote
  Sens.}, vol.~7, no.~6, pp. 1844--1868, 2014.

\bibitem{bioucas2012hyperspectral}
J.~M. Bioucas-Dias, A.~Plaza, N.~Dobigeon, M.~Parente, Q.~Du, P.~Gader, and
  J.~Chanussot, ``Hyperspectral unmixing overview: Geometrical, statistical,
  and sparse regression-based approaches,'' \emph{IEEE J. Sel. Top. Appl. Earth
  Observat. Remote Sens.}, vol.~5, no.~2, pp. 354--379, 2012.

\bibitem{dobigeon2009joint}
N.~Dobigeon, S.~Moussaoui, M.~Coulon, J.-Y. Tourneret, and A.~O. Hero, ``Joint
  bayesian endmember extraction and linear unmixing for hyperspectral
  imagery,'' \emph{IEEE Trans. Signal Process.}, vol.~57, no.~11, pp.
  4355--4368, 2009.

\bibitem{yang2015geometric}
S.~Yang, X.~Zhang, Y.~Yao, S.~Cheng, and L.~Jiao, ``Geometric nonnegative
  matrix factorization (gnmf) for hyperspectral unmixing,'' \emph{IEEE J. Sel.
  Top. Appl. Earth Observat. Remote Sens.}, vol.~8, no.~6, pp. 2696--2703,
  2015.

\bibitem{yuan2018overview}
J.~Yuan, Y.~Zhang, and F.~Gao, ``An overview on linear hyperspectral
  unmixing,'' \emph{J. Infrared Millim. Waves}, vol.~37, pp. 553--571, 2018.

\bibitem{drumetz2016blind}
L.~Drumetz, M.-A. Veganzones, S.~Henrot, R.~Phlypo, J.~Chanussot, and
  C.~Jutten, ``Blind hyperspectral unmixing using an extended linear mixing
  model to address spectral variability,'' \emph{IEEE Trans. Image Process.},
  vol.~25, no.~8, pp. 3890--3905, 2016.

\bibitem{hong2018augmented}
D.~Hong, N.~Yokoya, J.~Chanussot, and X.~X. Zhu, ``An augmented linear mixing
  model to address spectral variability for hyperspectral unmixing,''
  \emph{IEEE Trans. Image Process.}, vol.~28, no.~4, pp. 1923--1938, 2018.

\bibitem{chen2012nonlinear}
J.~Chen, C.~Richard, and P.~Honeine, ``Nonlinear unmixing of hyperspectral data
  based on a linear-mixture/nonlinear-fluctuation model,'' \emph{IEEE Trans.
  Signal Process.}, vol.~61, no.~2, pp. 480--492, 2012.

\bibitem{heinz2001fully}
D.~C. Heinz \emph{et~al.}, ``Fully constrained least squares linear spectral
  mixture analysis method for material quantification in hyperspectral
  imagery,'' \emph{IEEE Trans. Geosci. Remote Sens.}, vol.~39, no.~3, pp.
  529--545, 2001.

\bibitem{halimi2016fast}
A.~Halimi, J.~M. Bioucas-Dias, N.~Dobigeon, G.~S. Buller, and S.~McLaughlin,
  ``Fast hyperspectral unmixing in presence of nonlinearity or mismodeling
  effects,'' \emph{IEEE Trans. Comput. Imaging}, vol.~3, no.~2, pp. 146--159,
  2016.

\bibitem{ammanouil2015graph}
R.~Ammanouil, A.~Ferrari, and C.~Richard, ``A graph laplacian regularization
  for hyperspectral data unmixing,'' in \emph{2015 IEEE International
  Conference on Acoustics, Speech and Signal Processing (ICASSP)}.\hskip 1em
  plus 0.5em minus 0.4em\relax IEEE, 2015, pp. 1637--1641.

\bibitem{ammanouil2015hyperspectral_1}
------, ``Hyperspectral data unmixing with graph-based regularization,'' in
  \emph{2015 7th Workshop on Hyperspectral Image and Signal Processing:
  Evolution in Remote Sensing (WHISPERS)}.\hskip 1em plus 0.5em minus
  0.4em\relax IEEE, 2015, pp. 1--4.

\bibitem{iordache2012total}
M.-D. Iordache, J.~M. Bioucas-Dias, and A.~Plaza, ``Total variation spatial
  regularization for sparse hyperspectral unmixing,'' \emph{IEEE Trans. Geosci.
  Remote Sens.}, vol.~50, no.~11, pp. 4484--4502, 2012.

\bibitem{wang2017hyperspectral}
R.~Wang, H.-C. Li, A.~Pizurica, J.~Li, A.~Plaza, and W.~J. Emery,
  ``Hyperspectral unmixing using double reweighted sparse regression and total
  variation,'' \emph{IEEE Geosci. Remote Sens. Lett.}, vol.~14, no.~7, pp.
  1146--1150, 2017.

\bibitem{he2017total}
W.~He, H.~Zhang, and L.~Zhang, ``Total variation regularized reweighted sparse
  nonnegative matrix factorization for hyperspectral unmixing,'' \emph{IEEE
  Trans. Geosci. Remote Sens.}, vol.~55, no.~7, pp. 3909--3921, 2017.

\bibitem{feng2016nonlocal}
R.~Feng, Y.~Zhong, Y.~Wu, D.~He, X.~Xu, and L.~Zhang, ``Nonlocal total
  variation subpixel mapping for hyperspectral remote sensing imagery,''
  \emph{Remote Sensing}, vol.~8, no.~3, p. 250, 2016.

\bibitem{zhong2013non}
Y.~Zhong, R.~Feng, and L.~Zhang, ``Non-local sparse unmixing for hyperspectral
  remote sensing imagery,'' \emph{IEEE J. Sel. Top. Appl. Earth Observat.
  Remote Sens.}, vol.~7, no.~6, pp. 1889--1909, 2013.

\bibitem{yao2019nonconvex}
J.~Yao, D.~Meng, Q.~Zhao, W.~Cao, and Z.~Xu, ``Nonconvex-sparsity and
  nonlocal-smoothness-based blind hyperspectral unmixing,'' \emph{IEEE Trans.
  Image Process.}, vol.~28, no.~6, pp. 2991--3006, 2019.

\bibitem{wang2017spatial}
X.~Wang, Y.~Zhong, L.~Zhang, and Y.~Xu, ``Spatial group sparsity regularized
  nonnegative matrix factorization for hyperspectral unmixing,'' \emph{IEEE
  Trans. Geosci. Remote Sens.}, vol.~55, no.~11, pp. 6287--6304, 2017.

\bibitem{li2018superpixel}
Z.~Li, J.~Chen, and S.~Rahardja, ``Superpixel construction for hyperspectral
  unmixing,'' in \emph{2018 26th European Signal Processing Conference
  (EUSIPCO)}.\hskip 1em plus 0.5em minus 0.4em\relax IEEE, 2018, pp. 647--651.

\bibitem{Zhang2018Spectral}
S.~Zhang, J.~Li, H.-C. Li, C.~Deng, and A.~Plaza, ``Spectral--spatial weighted
  sparse regression for hyperspectral image unmixing,'' \emph{IEEE Trans.
  Geosci. Remote Sens.}, vol.~56, no.~6, pp. 3265--3276, 2018.

\bibitem{hong2018sulora}
D.~Hong and X.~X. Zhu, ``Sulora: Subspace unmixing with low-rank attribute
  embedding for hyperspectral data analysis,'' \emph{IEEE J. Sel. Top. Sig.
  Process.}, vol.~12, no.~6, pp. 1351--1363, 2018.

\bibitem{palsson2019convolutional}
B.~Palsson, M.~O. Ulfarsson, and J.~R. Sveinsson, ``Convolutional autoencoder
  for spatial-spectral hyperspectral unmixing,'' in \emph{2019 IEEE
  International Geoscience and Remote Sensing Symposium (IGARSS)}.\hskip 1em
  plus 0.5em minus 0.4em\relax IEEE, 2019, pp. 357--360.

\bibitem{khajehrayeni2020hyperspectral}
F.~Khajehrayeni and H.~Ghassemian, ``Hyperspectral unmixing using deep
  convolutional autoencoders in a supervised scenario,'' \emph{IEEE J. Sel.
  Top. Appl. Earth Observat. Remote Sens.}, vol.~13, pp. 567--576, 2020.

\bibitem{zhang2018hyperspectral}
X.~Zhang, Y.~Sun, J.~Zhang, P.~Wu, and L.~Jiao, ``Hyperspectral unmixing via
  deep convolutional neural networks,'' \emph{IEEE Geosci. Remote Sens. Lett.},
  vol.~15, no.~11, pp. 1755--1759, 2018.

\bibitem{lin2019hyperspectral}
B.~Lin, X.~Tao, and J.~Lu, ``Hyperspectral image denoising via matrix
  factorization and deep prior regularization,'' \emph{IEEE Trans. Image
  Process.}, vol.~29, pp. 565--578, 2019.

\bibitem{Zhuang2017Hyperspectral}
``Hyperspectral image denoising and anomaly detection based on low-rank and
  sparse representations,'' in \emph{Image and Signal Processing for Remote
  Sensing}, 2017.

\bibitem{gong2018learning}
D.~Gong, Z.~Zhang, Q.~Shi, A.~v.~d. Hengel, C.~Shen, and Y.~Zhang, ``Learning
  an optimizer for image deconvolution,'' \emph{arXiv preprint
  arXiv:1804.03368}, 2018.

\bibitem{wang2020learning}
X.~Wang, J.~Chen, C.~Richard, and D.~Brie, ``Learning spectral-spatial prior
  via {3DDnCNN} for hyperspectral image deconvolution,'' in \emph{2020 45th
  IEEE International Conference on Acoustics, Speech and Signal Processing
  (ICASSP)}.\hskip 1em plus 0.5em minus 0.4em\relax IEEE, 2020, pp. 2403--2407.

\bibitem{teodoro2017sharpening}
A.~Teodoro, J.~Bioucas-Dias, and M.~Figueiredo, ``Sharpening hyperspectral
  images using plug-and-play priors,'' in \emph{International Conference on
  Latent Variable Analysis and Signal Separation}.\hskip 1em plus 0.5em minus
  0.4em\relax Springer, 2017, pp. 392--402.

\bibitem{teodoro2017scene}
A.~M. Teodoro, J.~M. Bioucas-Dias, and M.~A. Figueiredo, ``Scene-adapted
  plug-and-play algorithm with convergence guarantees,'' in \emph{2017 IEEE
  27th International Workshop on Machine Learning for Signal Processing
  (MLSP)}.\hskip 1em plus 0.5em minus 0.4em\relax IEEE, 2017, pp. 1--6.

\bibitem{sreehari2016plug}
S.~Sreehari, S.~V. Venkatakrishnan, B.~Wohlberg, G.~T. Buzzard, L.~F. Drummy,
  J.~P. Simmons, and C.~A. Bouman, ``Plug-and-play priors for bright field
  electron tomography and sparse interpolation,'' \emph{IEEE Trans. Comput.
  Imaging}, vol.~2, no.~4, pp. 408--423, 2016.

\bibitem{teodoro2019image}
A.~M. Teodoro, J.~M. Bioucas-Dias, and M.~A. Figueiredo, ``Image restoration
  and reconstruction using targeted plug-and-play priors,'' \emph{IEEE Trans.
  Comput. Imaging}, vol.~5, no.~4, pp. 675--686, 2019.

\bibitem{boyd2011distributed}
S.~Boyd, N.~Parikh, E.~Chu, B.~Peleato, J.~Eckstein \emph{et~al.},
  ``Distributed optimization and statistical learning via the alternating
  direction method of multipliers,'' \emph{Found. Trends Mach. Learn.}, vol.~3,
  no.~1, pp. 1--122, 2011.

\bibitem{bai2020deep}
Y.~Bai, W.~Chen, J.~Chen, and W.~Guo, ``Deep learning methods for solving
  linear inverse problems: Research directions and paradigms,'' \emph{Signal
  Processing}, p. 107729, 2020.

\bibitem{buades2011non}
A.~Buades, B.~Coll, and J.-M. Morel, ``Non-local means denoising,'' \emph{Image
  Processing On Line}, vol.~1, pp. 208--212, 2011.

\bibitem{dabov2007image}
K.~Dabov, A.~Foi, V.~Katkovnik, and K.~Egiazarian, ``Image denoising by sparse
  3-d transform-domain collaborative filtering,'' \emph{IEEE Trans. Image
  Process.}, vol.~16, no.~8, pp. 2080--2095, 2007.

\bibitem{maggioni2012nonlocal}
M.~Maggioni, V.~Katkovnik, K.~Egiazarian, and A.~Foi, ``Nonlocal
  transform-domain filter for volumetric data denoising and reconstruction,''
  \emph{IEEE Trans. Image Process.}, vol.~22, no.~1, pp. 119--133, 2012.

\bibitem{wang2017hyperspectral_1}
Y.~Wang, J.~Peng, Q.~Zhao, Y.~Leung, X.-L. Zhao, and D.~Meng, ``Hyperspectral
  image restoration via total variation regularized low-rank tensor
  decomposition,'' \emph{IEEE J. Sel. Top. Appl. Earth Observat. Remote Sens.},
  vol.~11, no.~4, pp. 1227--1243, 2017.

\bibitem{zhang2017beyond}
K.~Zhang, W.~Zuo, Y.~Chen, D.~Meng, and L.~Zhang, ``Beyond a gaussian denoiser:
  Residual learning of deep cnn for image denoising,'' \emph{IEEE Trans. Image
  Process.}, vol.~26, no.~7, pp. 3142--3155, 2017.

\bibitem{shi2018collaborative}
Z.~Shi, T.~Shi, M.~Zhou, and X.~Xu, ``Collaborative sparse hyperspectral
  unmixing using $ {l_{0}} $ norm,'' \emph{IEEE Trans. Geosci. Remote Sens.},
  vol.~56, no.~9, pp. 5495--5508, 2018.

\bibitem{eckstein1992douglas}
J.~Eckstein and D.~P. Bertsekas, ``On the {Douglas--Rachford} splitting method
  and the proximal point algorithm for maximal monotone operators,''
  \emph{Mathematical Programming}, vol.~55, no. 1-3, pp. 293--318, 1992.

\bibitem{afonso2010fast}
M.~V. Afonso, J.~M. Bioucas-Dias, and M.~A. Figueiredo, ``Fast image recovery
  using variable splitting and constrained optimization,'' \emph{IEEE Trans.
  Image Process.}, vol.~19, no.~9, pp. 2345--2356, 2010.

\bibitem{aggarwal2016hyperspectral}
H.~K. Aggarwal and A.~Majumdar, ``Hyperspectral unmixing in the presence of
  mixed noise using joint-sparsity and total variation,'' \emph{IEEE J. Sel.
  Top. Appl. Earth Observat. Remote Sens.}, vol.~9, no.~9, pp. 4257--4266,
  2016.

\end{thebibliography}

\end{document}